\documentclass[prd,twocolumn,amsmath,nobibnotes,nofootinbib,superscriptaddress,preprintnumbers]{revtex4}

\usepackage{bm} \usepackage{graphicx}
\usepackage{epsfig}
\def\ba{\begin{eqnarray}}
\def\ea{\end{eqnarray}}
\def\be{\begin{equation}}
\def\ee{\end{equation}}
\def\({\left(}
\def\){\right)}
\def\[{\left[}
\def\]{\right]}
\def\<{\left<}
\def\>{\right>}

\newcommand{\rar}{\rightarrow}

\begin{document}

\title{Surviving the crash: assessing the aftermath of cosmic bubble collisions}
\date{\today}

\author{Anthony Aguirre}
\email{aguirre@scipp.ucsc.edu}
\affiliation{SCIPP, University of California, Santa Cruz, CA 95064, USA}
\author{Matthew C Johnson}
\email{mjohnson@theory.caltech.edu}
\affiliation{California Institute of Technology, Pasadena, CA 91125, USA}
\author{Martin Tysanner}
\email{tysanner@physics.ucsc.edu}
\affiliation{SCIPP, University of California, Santa Cruz, CA 95064, USA}

\begin{abstract}
This paper is the third in a series investigating the possibility that if we reside in an inflationary ``bubble universe", we might observe the effects of collisions with other such bubbles. Here, we study the interior structure of a bubble collision spacetime, focusing on the issue of where observers can reside. Numerical simulations indicate that if the inter-bubble domain wall accelerates away, infinite spacelike surfaces of homogeneity develop to the future of the collision; this strongly suggests that observers {\em can} have collisions to their past, and previous results then imply that this is very likely. However, for observers at nearly all locations, the restoration of homogeneity relegates any observable effects to a vanishingly small region on the sky. We find that bubble collisions may also play an important role in defining measures in inflation: a potentially infinite relative volume factor arises between two bubble types depending on the sign of the acceleration of the domain wall between them; this may in turn correlate with observables such as the scale or type of inflation.
\end{abstract}

\preprint{CALT-68.2707}

\maketitle

\section{Introduction}

In general models of cosmological inflation, the accelerated cosmic expansion is not a transient epoch, but rather continues forever and ends only {\em locally}, to create many `universes'\footnote{Meaning many radiation-dominated regions that would appear homogeneous and isotropic to observers within them.} with potentially different properties. For example, if the inflaton potential has a local minimum, inflation proceeds forever at the corresponding vacuum energy, but occasionally forms `bubbles' in which the field evolves to lower potential. This `false-vacuum eternal inflation' is not the only type: any region where the potential is sufficiently flat can also drive eternal inflation. (For recent reviews, see, e.g.,~\cite{Linde:2007fr,Aguirre:2007gy,Guth:2007ng,Vilenkin:2006ac}).

How seriously should we take these `other universes'? One might view them as `side effects' of a successful theory, but the features of an inflaton potential that yield successful observables are not inextricably tied to those driving eternal inflation. Another view is that although the predictions of cosmological properties would only be (at best) statistical in a diverse `multiverse', the statistics, along with observational selection effects, might leave hints of this fact in the values of, or correlations between, those observables. For example, a multiverse with many possible values of the cosmological constant, combined with `anthropic' constraints on which values allow observers, has gained notoriety as an explanation for why the observed cosmological constant has such an unnatural value from the standpoint of fundamental physics. But with few observables, a potentially very complex potential landscape, and other severe ambiguities, this program is both controversial and formidably difficult (see e.g.~\cite{Aguirre:2004qb,Aguirre:2005cj} for discussion).

Far better would be the possibility of directly observing some consequence of the `other universes.' One such opportunity arises due to the fact that during false vacuum eternal inflation, any given bubble universe will undergo an infinite number of collisions with other bubble universes. If our observable universe arose to the future of a collision (or many collisions), then perhaps the relics of such events remain to affect cosmological observables such as the CMB. This paper is the third in a series investigating this possibility. Along the way we will point out some findings regarding the dynamics of inflation and bubble collisions that may be important and useful even if bubble collisions are unobservable.

Determining the observable effects of bubble collisions presents a challenge. The collision dynamics specifies the structure of the bubble universe to its future, which in turn determines both the correct set of observers\footnote{By observers we mean geodesics emanating from some reheating surface inside of an `observation' bubble.} and what they will see. In~\cite{Aguirre:2007an}\footnote{This investigation was inspired by the pioneering study of~\cite{Garriga:2006hw}.} (hereafter Paper I) and~\cite{Aguirre:2007wm}\footnote{See~\cite{Chang:2007eq} for a similar study, and~\cite{Bousso:2006ge} for an earlier study of a special case.} (hereafter Paper II), we outlined the essential considerations in whether an observable bubble collision might be expected by us, and calculated the dynamics of bubble collisions in the limit of thin bubble walls and a thin domain wall separating the bubbles after a collision. However, these analytical models offered little insight into the structure of the bubble interior, and so it was assumed that observers follow geodesics associated with the original undisturbed bubble universe.

This assumption leads to some dramatic conclusions, as illustrated by the results of Paper I: all but a set of measure zero of observers would seem to be hit infinitely many, arbitrarily energetic and destructive, collisions that affect the whole sky! This is obviously inconsistent with the assumption that the bubble interior is unaffected, and might lead one to draw the {\em opposite} conclusion: that for a wide variety of bubble collisions, no (or few) observers even exist to the collision's future. Clearly, it is important to understand how including the effects of collisions on the bubble interior affects these conclusions. In Paper II, this question was briefly addressed, but in order to assemble a complete picture, it is necessary to employ numerical simulations of bubbles and their collisions, a task that we undertake in the present paper.

In Sec.~\ref{sec-bubnucevo} we review the nucleation and evolution of single vacuum bubbles. We then review the results of Paper I in Sec.~\ref{sec-probinfinity}, and argue in Sec.~\ref{sec-tameinfinity} how a reasonable definition of observers regulates the experienced energy of a collision. We describe our numerical simulations, and the relevant questions and assumptions in Sec.~\ref{sec-simulationssetup}. In Sec.~\ref{sec-singlebubblesim} and~\ref{sec-collsimulations} we present the results of our simulations for single and colliding bubbles. Extrapolating from the results of our simulations, we discuss the implications for observing bubble collisions in Sec.~\ref{sec-implications_obs} and for eternal inflation in Sec.~\ref{sec-implications_ei}. We conclude and discuss in Sec.~\ref{sec-discuss}. To avoid significant repetition of earlier work, we will sometimes refer to Papers I and II; we also adopt their notation.

\section{Bubble nucleation and evolution}\label{sec-bubnucevo}

The nucleation of a true vacuum bubble within a false vacuum de Sitter space is mediated by the instanton of lowest action which, when it exists, is the Coleman-DeLuccia (CDL) instanton~\cite{Coleman:1980aw}. The instanton is a field configuration on a generally compact, O(4) invariant Euclidean space
\begin{equation}
ds^2 = d\tau_E^2 + a(\tau_E)^2 d\Omega_3^2,
\end{equation}
where $d\Omega_3^2$ is the surface element of a unit 3-sphere. For a single minimally coupled scalar field $\phi$ with potential $V(\phi)$, the equations of motion for the field and scale factor are given by
\begin{eqnarray}\label{eq-CDLfield}
  \frac{dV}{d\phi} &=& \phi'' + \frac{3 a'}{a} \phi' ,
        \\ \label{eq-CDLscalef}
  a'^2 &=& 1 + \frac{8 \pi}{3} a^2 \left( \frac{\phi'^2}{2} - V \right),
\end{eqnarray}
where the primes are derivatives with respect to $\tau_E$. One must solve these coupled equations for nonÐsingular solutions that travel between two field endpoints with vanishing velocity.  We refer readers to the large existing literature on the CDL instanton for more details on its construction (see e.g.~\cite{Banks:2002nm}).

\begin{figure*}[htb]
\includegraphics[width=14cm]{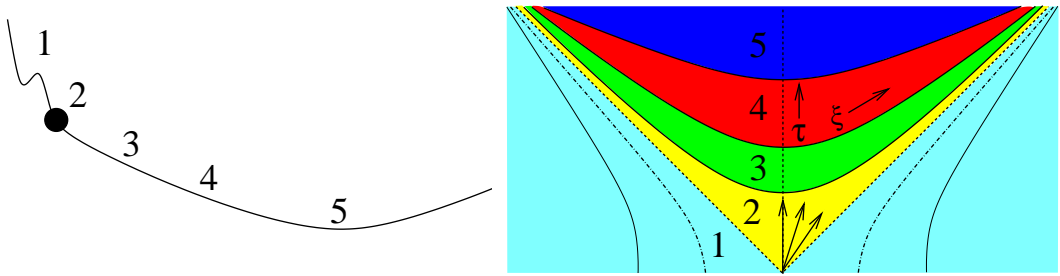}
\caption{The relation between the scalar potential and the bubble spacetime. The open patch with metric~\eqref{eq-frwpatch} fills the forward lightcone emanating from the nucleation center, where $\tau$ labels successive spacelike hyperboloids and $\xi$ is the coordinate distance from the origin on each hyperboloid. Constant-$\xi$ trajectories can also be viewed as geodesics passing through the nucleation center with various boosts, as indicated by the congruence at the bottom of the picture. Region 1, which comprises the bubble wall, is determined by the analytic continuation of the instanton, whose endpoint (black dot on the potential) analytically continues to the light cone (dashed line between regions 1 and 2). Region 2 is the period of curvature domination, which then gives way to inflation in region 3 as the universe expands. Inflation ends in region 4, and the field oscillates and reheats the universe. In region 5, the field settles into its minimum and standard big bang FLRW cosmology begins.
  \label{fig-cartoon}
}
\end{figure*}

Analytically continuing, one obtains initial data for the subsequent Lorentzian evolution of the bubble interior and exterior. Because of the O(4)-invariance of the instanton, the post-nucleation spacetime will possess SO(3,1) symmetry, and include a causally complete open Friedmann-Lemaitre-Robertson-Walker (FLRW) universe filling the forward light cone emanating from the nucleation center. Extending the coordinates across the light cone, one enters the wall of the bubble, in which the field and geometry interpolate between the endpoints of the instanton on the true and false vacuum sides of the potential barrier. A cartoon of the relation between the scalar potential and the postÐnucleation space time is shown in Fig.~\ref{fig-cartoon}.

The FLRW patch inside of the bubble has the `open slicing' metric
\begin{equation}\label{eq-frwpatch}
ds^2 = -d\tau^2 + a(\tau)^2 \left[ d \xi^2 + \sinh^2 \xi \, d\Omega_2^2 \right],
\end{equation}
where $0 \leq \tau < \infty$ and $0 \leq \xi < \infty$ (the Milne slicing of Minkowski space is obtained by setting $a(\tau) = \tau$). The field is homogeneous on the null $\tau = 0$ surface, with zero kinetic energy and a field value determined by the instanton endpoint. Subsequent equal-$\tau$ surfaces are homogeneous spacelike hyperboloids and coincide with surfaces of constant $\phi$. The hypersurface orthogonal geodesics $\xi = {\rm const.}$ all experience the same isotropic and homogeneous cosmological evolution, rendering this the preferred congruence to associate with observers inside of the bubble. The members of this congruence can be labeled by a relative boost with respect to the nucleation center as depicted in Fig.~\ref{fig-cartoon}, with the boost diverging as $\xi \rightarrow \infty$.

In region 2 of Fig.~\ref{fig-cartoon} (small $\tau$), curvature dominates the energy density, and the scale factor evolves like $a \propto \tau$. As $\tau$ increases, the field $\phi$ evolves from the tunneled-to value, through a (presumed) inflationary phase in region 3 of Fig.~\ref{fig-cartoon}, into reheating in region 4, and finally entering standard big-bang cosmological evolution in region 5. Models of this type go by the name of open inflation, and there is a large body of literature concerning their phenomenology (see e.g.~\cite{GarciaBellido:1997uh} and references therein). To preserve the continuity of discussion, we postpone further discussion of model building in open inflation to Appendix~\ref{sec-potentials}, where we present a number of new results and explicit potentials.

\section{Problems with infinities in bubble collisions}\label{sec-probinfinity}

If a bubble in de Sitter space nucleates inside the Hubble volume $V_4 \sim H_F^{-4}$ of another bubble, then the two will eventually collide. Focusing on an 'observation' bubble, consider observers at some radius $\xi_\mathrm{obs}$ at a time $\tau_\mathrm{obs}$.  Under the assumption that the structure of the interior of the observation bubble is unaffected by collisions, it was shown in~\cite{Garriga:2006hw} and elaborated on in Paper I that observers at $(\tau_\mathrm{obs}, \xi_\mathrm{obs})$ will see bubbles enter their past lightcone at a rate that depends on $\tau_\mathrm{obs},\ \xi_\mathrm{obs}$ and the direction on the sky.  The area on the sky affected by other bubbles is a set of disks with an angular size distribution that depends on these same variables.  (We define the sky of the observer as a two-sphere formed by the intersection of the observer's past lightcone and a surface of constant $\tau < \tau_\mathrm{obs}$.)  The lowest collision rate occurs at the `center' of the bubble at $\xi_\mathrm{obs} = 0$; this spatial origin is at rest in a preferred frame of the background space, i.e. a remnant of the initial condition surface imposed long before the bubble's formation~\cite{Garriga:2006hw}.

Now a constant-time surface of the observation bubble without collisions is homogeneous, so there is a natural measure on observers by physical volume, $dV_3 = a^3 \sinh \xi \sin \theta d\xi d\theta d\phi$. Thus, neglecting collisions, all observers should be at $\xi \rightarrow \infty$ (where `all' means all but the set of measure zero that are within any given radius $\xi$).

As shown in~\cite{Garriga:2006hw} and Paper I, these observers formally have infinitely many bubbles in their past lightcone. This occurs because these observers have worldlines that look null with respect to the background preferred frame (see Fig.~\ref{fig-cartoon} at small $\tau$), and the collision rate is therefore modified by an infinite time dilation factor. Paper I also showed that each collision affects essentially the full sky of the observer.

Further, these collisions naively appear to be completely catastrophic. One way to consider this would be to note that to a mote of dust at rest in the background frame, a $\xi \rightarrow \infty$ observer would appear to be null, i.e. have an infinite collision energy. The same would seem to be true of a  bubble collision (except that it would be even worse, with the bubble wall accelerating toward the observer.) Another way to view the situation is that a global 'boost' can be performed on the spacetime to move the observer to $\xi_{\rm obs} = 0$ (see Papers I and II for details). This in turn moves the colliding bubble back toward past null infinity, seemingly giving it an infinite time to accelerate toward the observation bubble and making it extremely dangerous.

How can we make sense of these infinitely many all-sky infinite-energy collisions? Can observers actually exist in the future lightcone of these collisions and see them? If so, then arguably `all' observers could have collisions to their past.

\section{Taming infinities in collisions}\label{sec-tameinfinity}

Even without addressing the effects of collisions on the bubble interior, we can make an important distinction about what types of observers to consider.  Agreement with modern cosmological observations requires a nearly homogeneous and isotropic reheating surface onto which we can match the standard big bang cosmological evolution.  Anything that occurs prior to the reheating time $\tau_\textrm{\tiny{RH}}$ merely sets the initial conditions for the cosmology we observe.  Hence we define an observer to mean a member of a hypersurface-orthogonal geodesic congruence emanating from the reheating surface.  In the limit where no collisions disturb the bubble interior, this definition suggests we should restrict our attention to events at times $\tau > \tau_\textrm{\tiny{RH}}$.  A restriction to finite $\tau$, as we will see, regulates the collision energy that a member of the $\xi = \mathrm{const.}$ congruence will observe.

Consider the limit where no field evolution occurs inside the observation bubble, so that the cosmological constant and Hubble parameter $H$ inside are approximately equal to those of the exterior false vacuum: $H \simeq H_F$.  Then we can use any of the standard slicings of de Sitter space (open, closed, or flat) to  coordinatize the bubble interior and ask:  what is the relative velocity between an observer at fixed $(\tau_\mathrm{obs},\xi_\mathrm{obs})$ and a curve of constant acceleration?  (The relative velocity offers an analogue for the kinetic energy of an incoming bubble wall that an observer would record.)

\begin{figure} [!h]
 \begin{center}
   \includegraphics[width=3.25in]{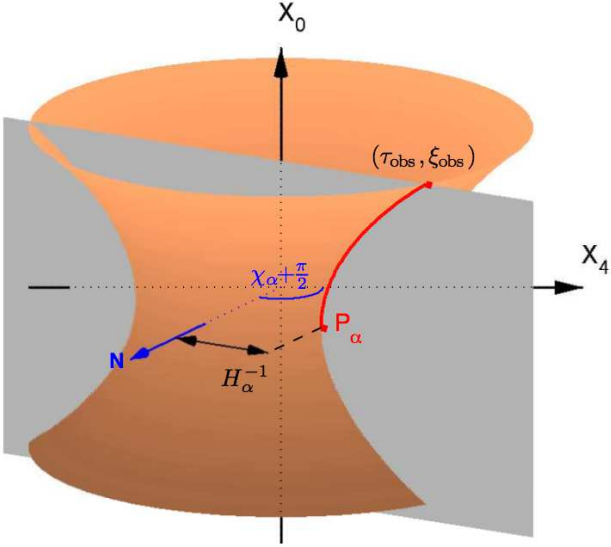}
 \end{center}
 \caption{\label{FigSlicedHyperboloid}
 An accelerated world line in the background de Sitter spacetime (solid red curve) starts at rest at a nucleation point $P_\alpha$ and meets an observer at some $(\tau_\mathrm{obs},\xi_\mathrm{obs})$ in an open universe that also nucleates at $X_0 = 0$.  The trajectory lies on the intersection of the hyperboloid and a hyperplane that is a distance $\alpha$ from the $X_0$ axis and parallel to it.  The angle $\chi_\alpha$ between the plane and the $X_4$ axis is fixed by the observer position and $\alpha$.}
 \vspace{-4pt}
\end{figure}

Specifically, given the dS hyperboloid $X^\mu X_\mu = H^{-2}$ embedded in 5-D Minkowski space, consider a one-parameter family of accelerated worldlines $X_\alpha = X(\mathcal{T}; \alpha)$.  An observation bubble, coordinatized by the open slicing of dS, nucleates at global time $\mathcal{T} = X_0 = 0$.  Each accelerated worldline $X_\alpha$ can be thought of as the wall of a different bubble that also nucleates at $X_0=0$ with initial velocity $v=d X_\alpha/\/d X_0 = 0$ at some characteristic point $P_\alpha$ on the hyperboloid; this `collides' at open slicing time $\tau_\mathrm{obs}$ with an observer at rest inside the observation bubble, i.e. on the worldline $x_\mathrm{obs} = (\tau, \xi_\mathrm{obs}, 0, 0)$, $\xi_\mathrm{obs}$ constant.  (Within this setup we are working in the `collision frame' of Paper~II, so that choosing the common nucleation time $X_0 = 0$ does not sacrifice generality.)  Then $P_\alpha$ is fixed by $(\tau_\mathrm{obs}, \xi_\mathrm{obs})$ and the desired acceleration $\ddot X_\alpha(\mathcal{T})$.  The problem is to kinematically evolve the wall trajectory $X_\alpha(\mathcal{T})$ into the observation bubble and compute the relative boost $\gamma_\alpha(\tau_\mathrm{obs}, \xi_\mathrm{obs})$; see Fig.~\ref{FigSlicedHyperboloid}.

Acceleration curves having the desired initial condition $v = 0$ at $X_0 = 0$ can be obtained by slicing the dS hyperboloid with a hyperplane that is parallel to and a distance $\alpha$ from the $X_0$ axis.  In the limit $\alpha = H^{-1}$, the hyperplane is tangent to the hyperboloid at $X_0=0$ and the curve is null; in the limit $\alpha = 0$, the hyperplane passes through the origin, and the curve is a geodesic.  In between should lie all accelerations.  The trick is to find the hyperplane that contains both the point $(\tau_\mathrm{obs}, \xi_\mathrm{obs})$ inside the bubble and the desired accelerated worldline $X_\alpha(t_\alpha)$, determine the hyperbolic (constant acceleration) trajectory in suitable coordinates $(t_\alpha,\chi_\alpha)$ on the hyperplane, and finally transform $X_\alpha$ from those coordinates to the open slicing at $(\tau_\mathrm{obs}, \xi_\mathrm{obs})$ and compute the boost
\begin{equation}  \label{EqnBoostPrescription}
 g_{\mu\nu} U_\mathrm{\alpha}^\mu \, U_\mathrm{obs}^\nu
    = - \frac{d\tau}{d t_\alpha}
    \equiv - \gamma_\alpha
 \,.
\end{equation}

Fig.~\ref{FigSlicedHyperboloid} shows the basic setup.  We fix the angular coordinates at $\theta = \pi/\/2$ and $\phi = 0$ and consider only $(X_0,X_1,X_4)$ with no loss of generality.  The desired hyperplane that contains both $X_\alpha(t_\alpha)$ and $(\tau_\mathrm{obs}, \xi_\mathrm{obs})$ satisfies \mbox{$X_1 \cos\chi_\alpha - X_4 \sin\chi_\alpha = \alpha$}, $0 \le \alpha < H^{-1}$.  Then $\chi_\alpha$ measures the angle between the hyperplane and the $X_4$ axis (compare with $\chi$ in the closed slicing).

 The curve $X_\alpha(t_\alpha)$ is the hyperboloid given by
\begin{equation}  \label{EqnAccelerationEqn}
 -X_0^2 + (X_1-X_1^\prime)^2 + (X_4-X_4^\prime)^2 \; = \; H_\alpha^{-2}
 \; \equiv \; H^{-2} - \alpha^2
 \,.
\end{equation}
That is, its origin is the intersection of the hyperplane and the normal $N$ that also passes through the embedding space origin:  $X^\prime = (0, \alpha\cos\chi_\alpha, -\alpha\sin\chi_\alpha)$.   Eq.~\eqref{EqnAccelerationEqn} has the embedding
\begin{eqnarray}  \label{EqnAccelCurveEmbedding}
 X_0 &=& H_\alpha^{-1} \sinh(H_\alpha t_\alpha)  \nonumber \\
 X_1 &=& H_\alpha^{-1} \cosh(H_\alpha t_\alpha) \sin\chi_\alpha +
               \alpha\cos\chi_\alpha \\
 X_4 &=& H_\alpha^{-1} \cosh(H_\alpha t_\alpha) \cos\chi_\alpha -
               \alpha\sin\chi_\alpha  \nonumber
 \,.
\end{eqnarray}
This becomes the closed chart when $\alpha = 0$ so that $X_\alpha(t_\alpha)$ is a geodesic.

Comparing the embedding Eqs.~\eqref{EqnAccelCurveEmbedding} to the embedding of the open chart (see, e.g. Paper~I), the specific choice of $\alpha$ and observer position $(\tau_\mathrm{obs},\xi_\mathrm{obs})$ fix the angle to a characteristic constant value,
\begin{multline}  \label{EqnPlaneAngle}
\hspace{-10pt}
 \cos\chi_\alpha
   = \frac{ \alpha H\sinh H\tau_\mathrm{obs} \sinh\xi_\mathrm{obs} }
          { \sinh^2 H\tau_\mathrm{obs} \cosh^2\xi_\mathrm{obs} + 1 }
          \\ \quad
     + \; \frac{ \cosh H\tau_\mathrm{obs}
            \left(
               \sinh^2 H\tau_\mathrm{obs} \cosh^2\xi_\mathrm{obs}
               + 1 - \alpha^2 H^2
            \right)^{1/\/2} }
          { \sinh^2 H\tau_\mathrm{obs} \cosh^2\xi_\mathrm{obs} + 1 }
 \,.
\end{multline}

The boost $\gamma_\alpha \equiv d\tau/\/d t_\alpha$ follows from  Eq.~\eqref{EqnPlaneAngle} and the embeddings for $X_0$ and $X_4$ in the same charts.  At $(\tau_\mathrm{obs}, \xi_\mathrm{obs})$,
\begin{multline}  \label{EqnBoost}
\hspace{-6pt}
 \gamma_\alpha
   = \frac{\cosh H\tau_\mathrm{obs}} { \sqrt{1 - \alpha^2 H^2} }
     \Bigg[
       \frac{ \alpha H\tanh H\tau_\mathrm{obs} \tanh\xi_\mathrm{obs}}
            { \cosh^2 H\tau_\mathrm{obs} - \tanh^2\xi_\mathrm{obs} }  \\
     + \; \frac{ \left(
           \cosh^2 H\tau_\mathrm{obs}
             - \tanh^2\xi_\mathrm{obs}
             - \frac{\alpha^2 H^2}{\cosh^2\xi_\mathrm{obs}}
        \right)^{1/\/2} }
            { \cosh^2 H\tau_\mathrm{obs} - \tanh^2\xi_\mathrm{obs} }
     \Bigg]
 \,.
\end{multline}
The characteristic `Hubble parameter' $H_\alpha$ has been eliminated by $H^2 H_\alpha^{-2} = 1 - \alpha^2 H^2$.

When $\alpha = 0$, expression \eqref{EqnBoost} is the boost between the observer at $(\tau_\mathrm{obs},\xi_\mathrm{obs})$ in the open slicing and an observer at rest at the \textit{same} spacetime point in the closed slicing, i.e.
\begin{equation}  \label{EqnBoostSameH}
 \gamma_0 (\tau_\mathrm{obs},\xi_\mathrm{obs})
   = \left(
           1 - \frac{ \tanh^2{\xi_\mathrm{obs}} }
                    { \cosh^2\! H\tau_\mathrm{obs} }
    \right)^{-1/\/2}
 \,,
\end{equation}
which is equivalent to the expression of~\cite{Garriga:2006hw}.

Consider the general boost~\eqref{EqnBoost} in certain limits.  For $\alpha \rightarrow H^{-1}$, the acceleration curve becomes null for all $\tau_\mathrm{obs}$ and $\xi_\mathrm{obs}$, as it should because $\alpha = H^{-1}$ generates the lightcone.  At late times,
\begin{equation}  \label{EqnLateTimeBoost}
 \lim_{\tau_\mathrm{obs} \rightarrow \infty} \gamma_\alpha
   = (1 - \alpha^2 H^2)^{-1/\/2}
 \,;
\end{equation}
the boost ``saturates'' to the same value for all $\xi_\mathrm{obs}$.  Finally, at large radii,
\begin{equation}  \label{EqnInfiniteXi}
 \lim_{\xi_\mathrm{obs} \rightarrow \infty} \gamma_\alpha
   = \frac{\coth H\tau_\mathrm{obs}} {\sqrt{1 - \alpha^2 H^2}}
       \left( \frac{\alpha H}{\cosh H\tau_\mathrm{obs}} + 1 \right),
\end{equation}
which is finite for any given $\tau_\mathrm{obs}$ (though as $\tau_\mathrm{obs} \rightarrow 0$, the boost is arbitrarily large).  Thus we see that, in this model, restricting to $\tau_\mathrm{obs} > 0$ regulates the infinite relative boost between an observer and an incoming bubble.

We now turn to a more detailed, numerical study of the bubble interior that includes the effects of collisions. This will allow us to address the question of observers and what they see in more detail.

\section{Simulations}\label{sec-simulationssetup}

As we just saw, the effects of a collision drastically depend on the vantage point.  Thus, in assessing observational signatures of bubble collisions, the question of {\em where the observers are} is central. Previously studied analytic~\cite{Aguirre:2007wm,Chang:2007eq} models can only go so far in this regard, offering insight into the collision itself but saying nothing about the distribution or existence of observers in its aftermath. This distribution is determined by the surfaces of homogeneity inside the bubble.

To study the structure of these surfaces, we turn to a numerical analysis in this section. In particular, we address the following set of questions
\begin{itemize}
\item Are there still (approximately) homogeneous infinite spatial slices foliating the region to the future of a bubble collision event, as hypothesized in paper II?
\item Can inflation occur to the future of a collision, and thus produce a reheating surface that is very nearly homogeneous, flat, and isotropic?
\item Associating observers with such a reheating surface, are the observable properties of a collision different than those seen by observers defined with respect to the geodesic congruence of an undisturbed bubble?
\end{itemize}
To address these questions, we will find it helpful to consider both collisions between SO(3,1)-symmetric bubbles and O(3)-symmetric bubbles in isolation (which will yield more additional information about evolution from inhomogenous initial data).

\subsection{Method}\label{sec-method}

The SO(2,1) hyperbolic symmetry of the collision region between two bubbles, and O(3) invariance of spherically symmetric single bubbles, allow us to reduce the problem to 1+1 dimensions, rendering a simulation computationally inexpensive.\footnote{Such calculations have previously been carried out in~\cite{Hawking:1982ga,Kosowsky:1991ua,Blanco-Pillado:2003hq}.  The SO(2,1) is the residual symmetry remaining after the SO(3,1) symmetry of the individual bubbles is broken.  In Minkowski space this breaking can be thought of as the choice of the particular frame in which they both nucleate at the same time.} A full treatment would include additional constituents coupled to the inflaton field, and gravitational effects.  Here, we simulate the field dynamics alone, in flat spacetime.  Accordingly, we focus on cases where gravitational effects are unimportant, and assume that other fields only become relevant during the reheating epoch; the qualitative answers provided by these cases should not change with the inclusion of gravitational effects.

In the spherical case with standard coordinates $(t,r,\theta,\varphi)$, the field $\phi(r,t)$ is governed by
\begin{equation}\label{eq-O3evolve}
\partial_t^2\phi-\partial_r^2\phi-{2\over r}\,\partial_r\phi
  = -{dV\over d\phi}.
\end{equation}
Each point in the $(t,r)$ simulation plane corresponds to a two-sphere of radius $r$.

\begin{figure}[htb]
\includegraphics[width=7.5cm,angle=90]{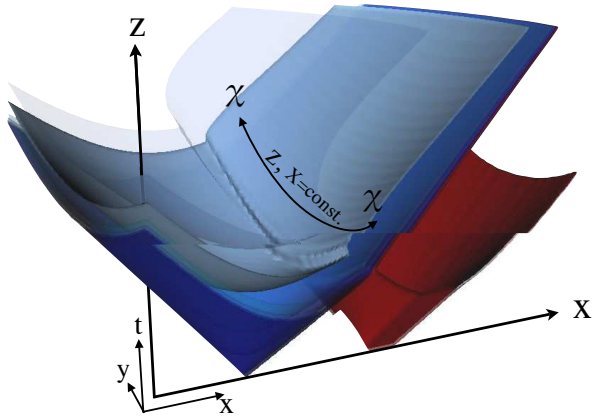}
\caption{
A 2+1D slice of the collision between two bubbles in Minkowski space (plotting several constant-field surfaces from the fiducial simulation below) showing the relation between the Cartesian coordinates $(t,x,y,w)$ (with $w=0$ suppressed) and the hyperbolic coordinates $(z,x,\chi,\varphi=0)$. The depicted lines of constant $z$ are hyperbolae in the Cartesian coordinates. Given the SO(2,1) symmetry, the simulation region represents the region above the planes $y=\pm t$, or can alternatively be taken as the slice $(y=w=0)$, i.e. the Cartesian $t-x$ plane.
  \label{fig-minkcoll}
}
\end{figure}

The hyperbolic case has a flat-space metric (see Ref.~\cite{Wu:1984eda}),
\begin{equation}\label{eq-xzmetric}
ds^2 = -dz^2 + dx^2 + z^2 (d \chi^2 + \sinh^2 \chi d\varphi^2),
\end{equation}
and $\phi(z,x)$ obeys
\begin{equation}\label{eq-hypevolve}
\partial_z^2\phi-\partial_x^2\phi+{2\over z}\,\partial_z\phi
  = -{dV\over d\phi}.
\end{equation}
The relation to the flat-space Cartesian coordinates is given by
\begin{eqnarray}  \label{eq-HyperbolicCoordTransform}
&& t = z \cosh \chi,\qquad\quad\,  x = x, \\ \nonumber
&& y = z \sinh \chi \cos \varphi, \ \  w = z \sinh \chi \sin \varphi.
\end{eqnarray}
In this case, each point in the $(z,x)$ simulation plane will correspond to a two-hyperbola of radius $z$.
A depiction of the collision region, and the relation between the Cartesian and hyperbolic coordinates, is shown in Fig.~\ref{fig-minkcoll} (we refer the reader to Paper II for further details).

For a potential of the form $V=\mu^4 v(\phi/M)$, where $\mu$ and $M$ are mass scales, we can rescale the variables,
\begin{equation}\label{eq-dimless}
\phi/M \rar \phi,\ \ {\mu^2\over M}r \rar r,\ \  {\mu^2\over M}t \rar t,\ \ \mu^4 V \rar V,
\end{equation}
so that $\phi,r,t,$ (or $\phi,x,z$) and $V$ are dimensionless; plotted results will be shown in these terms. The numerical implementation is as in Ref.~\cite{Hawking:1982ga}, using a regular diamond grid of spacing $2h$ with equations similar to Eqs. (21), (22), and (23) of that paper.
In the spherical case, the boundary at $r=0$ is treated by requiring $\partial_r\phi=0$. In the hyperbolic case, and for large-$r$ in the spherical case, we use reflecting boundary conditions, but exclude regions where field values differing significantly from the background false vacuum are reflected. We start the calculation at $t=0$ or $z=0$, using initial bubble profiles $\phi(r)$ or $\phi(x)$ calculated numerically by solving the instanton equations, as described below, with $\partial_t \phi=0$ or $\partial_z \phi=0$ respectively. The algorithm is stable, and we have checked that none of our results depend on the resolution chosen.

\begin{figure}[htb]
\includegraphics[width=7cm,angle=90]{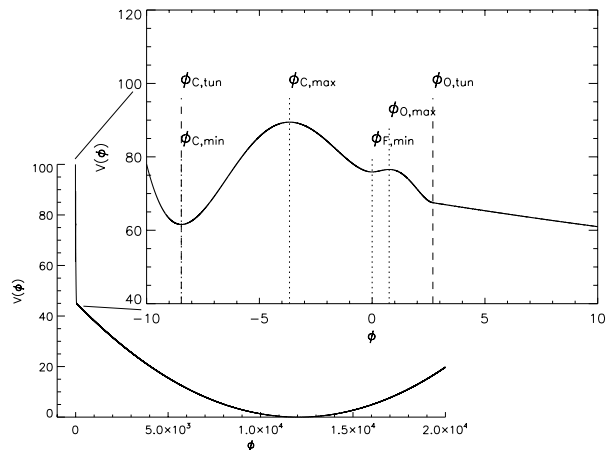}
\caption{
The re-scaled large--field potential used in numerically evolving $\phi$. Important field values are labeled with a `C' or `O' superscript for the collision or observation bubble, a `min' or `max' superscript for local minima or maxima, and a `tun' superscript for values of the (post-tunneling) instanton endpoint.  A selection of these values are also labeled in the plots of $\phi(x,z)$ and $\phi(r,t)$.
  \label{fig-simpotential}
}
\end{figure}

\begin{figure}[htb]
\includegraphics[width=8.5cm]{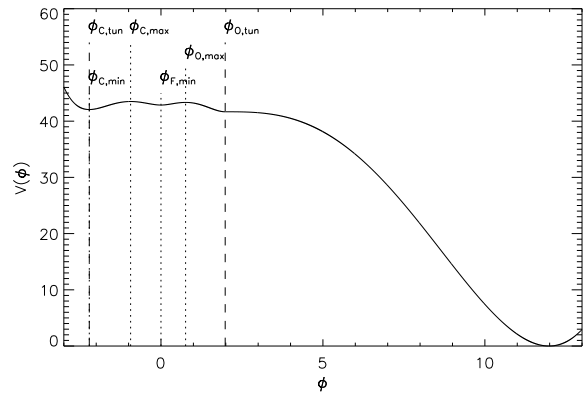}
\caption{
The re-scaled small--field potential used in numerically evolving $\phi$. Important field values are labeled with a `C' or `O' superscript for the collision or observation bubble, a `min' or `max' superscript for local minima or maxima, and a `tun' superscript for values of the (post-tunneling) instanton endpoint.
  \label{fig-simpotential2}
}
\end{figure}

To make connection with realistic models of open inflation, we have constructed potentials that, were gravitational effects included, would produce a phenomenologically acceptable epoch of inflation inside of the observation bubble. Appendix~\ref{sec-potentials} describes their construction. We use the four-segment large--field piecewise potentials depicted in Fig.~\ref{fig-simpotential} and the small--field potential depicted in Fig.\ref{fig-simpotential2} throughout the rest of the paper.

We will focus on the large--field model as our fiducial potential for most examples, and we use the parameters (in units where the Planck mass $M_p = 1$) $\{\mu_1^4 = 1.5 \times 10^{-13},\  \mu_2^4 = 1 \times 10^{-13}, \ M_1 = 7.4 \times 10^{-4},\ M_2 = 2.5 \times 10^{-4},\ M_4 = 10^{-6} \}$.  We re-scale variables as in Eq.~\eqref{eq-dimless} using $\mu = \mu_2$ and $M = M_2$, which are the scales controlling the barrier associated with the observation bubble. Fig.~\ref{fig-simpotential} shows the re-scaled potential. We will also have occasion to use the small--field potential depicted in Fig.~\ref{fig-simpotential2}, which is re--scaled similarly to the large--field example.

In keeping with our neglect of gravitational effects, it is important to ensure that the potential gives rise to bubbles whose initial radius is much smaller than the false vacuum horizon size. In this case, the CDL equations of motion~\eqref{eq-CDLfield} and \eqref{eq-CDLscalef} reduce to $a \propto \tau_E$ and
\begin{equation}\label{eq-CDLinitialdata}
\partial_{\tau_E}^2 \phi + \frac{3}{\tau_E} \partial_{\tau_E} \phi= \frac{dV}{d\phi} \,,
\end{equation}
with $\tau_E \equiv \sqrt{r^2+t^2}$~\cite{Coleman:1977py}. We numerically solve this equation of motion for solutions that have one instanton endpoint at $\tau_E = 0$ in the basin of attraction of the true vacuum (where $\partial_{\tau_E} \phi = 0$), and the other at the false vacuum as $\tau_E \rightarrow \infty$.

When we assume hyperbolic symmetry, we can use the instanton profile $\phi(\tau_E)$ to describe the initial configuration $\phi(x;z=0)$ because $x=\tau_E$ along the line $z=\chi=\phi = 0$. Evaluating the initial radius for the potential shown in Fig~\ref{fig-simpotential}, we find that $H R_0 \simeq 6 \times 10^{-2}$ for both bubble types; see Eq.~\eqref{eq-deSitterHorizonSize}. To evolve configurations with more than one bubble, we simply glue together the bubbles' initial conditions: for sufficiently widely-separated nucleations this introduces negligible error as both configurations are exponentially close to the false vacuum far from the nucleation center.

In addition to the CDL solutions, we will also consider configurations close to the to the O(3)-invariant static bubbles, given by the solution to
\begin{equation}\label{eq-O3initial}
\partial_{\tau_E}^2 \phi + \frac{2}{\tau_E} \, \partial_{\tau_E} \phi= \frac{dV}{d\phi} \,.
\end{equation}
It has boundary conditions identical to the O(4) symmetric case discussed above.

\begin{figure}[htb]
\includegraphics[width=8.5cm]{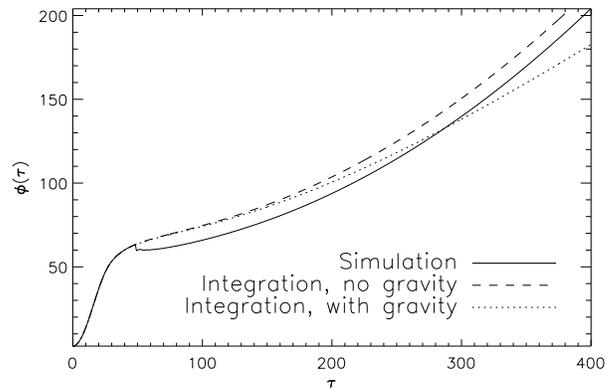}
\caption{Evolution of $\phi(\tau)$ for a direct integration of the field equation for homogeneous slices of constant $\tau$ with (dotted line) or without (dashed line) gravity.  The solid line is the simulation field at $\xi=0$; the corner at $\tau\sim 50$ is the impact of the colliding bubble.
  \label{fig-gandnog}
}
\end{figure}

The energy density on constant-$\phi$ slices inside of a CDL bubble is dominated by the effective energy of 3-curvature for some range of $\tau$ after the bubble nucleation, as discussed in Sec.~\ref{sec-bubnucevo}. In this regime, $a \propto \tau$, and the gravitational and non-gravitational field equations are identical. At some point in the evolution, the potential energy of the field becomes important; this marks the beginning of the inflationary epoch inside of the bubble, and the gravitational and non-gravitational solutions will deviate. Fig.~\ref{fig-gandnog} shows the field evolution $\phi (\tau)$ for our example potential in these two cases, in terms of the open-slicing coordinates~\eqref{eq-frwpatch}. It can be seen that the non-gravitational field evolution is a good approximation to the true evolution for approximately 200 simulation time units, of order 1-2 false-vacuum Hubble times.

Since we will consider deviations from the SO(3,1) symmetry of a CDL bubble, we need a general, local definition of the conditions necessary to cause inflation. We define an inflating region as satisfying two criteria, as in Ref.~\cite{Aguirre:2007gy}. First, we require that geodesic congruences undergo accelerated expansion. This requires a violation of the Strong Energy Condition (SEC): in Appendix~\ref{sec-SEC}, we derive a convenient form of the SEC in the presence of a general scalar field configuration. Second, we require that such congruences eventually intersect a relatively homogeneous reheating surface. A spacetime that meets both of these conditions {\em somewhere} can support inflation. The phenomenological viability of any given realization of inflation, of course, is a detailed question that must be evaluated on a case-by-case basis.

We now present the results of our simulations, first for single bubbles and then for bubble collisions.

\subsection{Single-bubble simulations}\label{sec-singlebubblesim}

We have applied our simulation to the evolution of both SO(3,1) and O(3)-invariant single bubbles. The initial data for the O(3,1)-invariant case comes from solving Eq.~\eqref{eq-CDLinitialdata} and then evolving with Eq.~\eqref{eq-O3evolve}, as described above. The SO(3,1)-invariant  observation bubble for our fiducial potential is shown in Fig.~\ref{fig-O(3,1)singlebubble}. Filling the light cone emanating from the nucleation center are the spacelike surfaces of constant field that evolve as the field rolls down the slow-roll region of the potential towards the true vacuum. As discussed above, the simulation results neglecting gravity are identical to the results obtained when including gravity for some range of $\tau$ inside of the bubble; more generally, flat-space is a good approximation within a fixed invariant distance $\sim H^{-1}\approx 160$ from the origin that includes all of the region shown in Fig.~\ref{fig-O(3,1)singlebubble}.

\begin{figure*}[htb]
\includegraphics[width=12.cm]{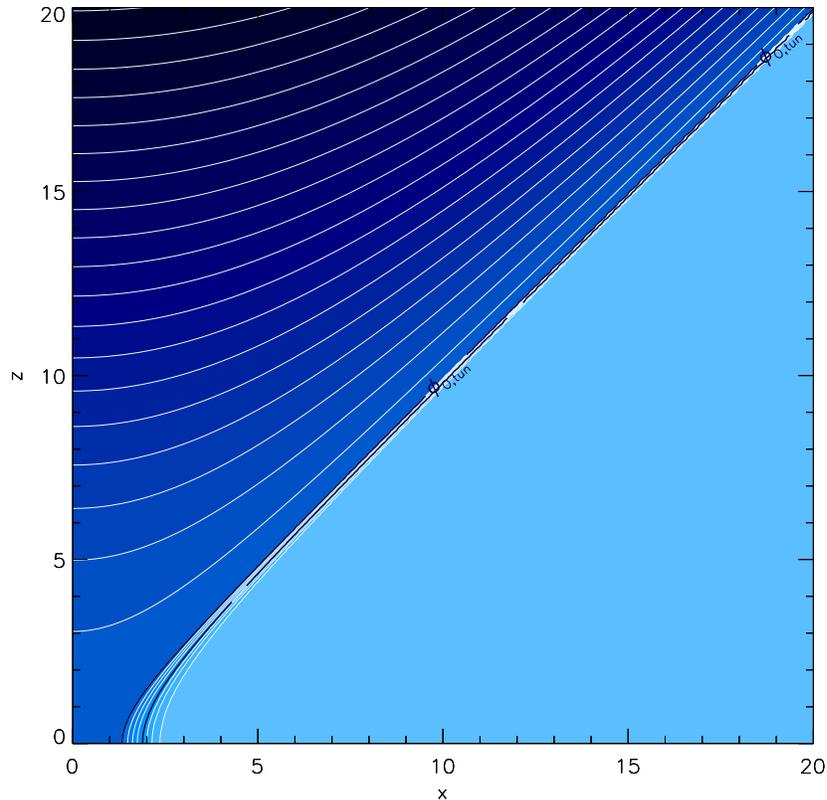}
\caption{An SO(3,1) invariant bubble solution. The spacelike surfaces of homogeneity foliate the interior of the light cone emanating from the nucleation center.
  \label{fig-O(3,1)singlebubble}
}
\end{figure*}

\begin{figure*}[htb]
\includegraphics[width=12.cm]{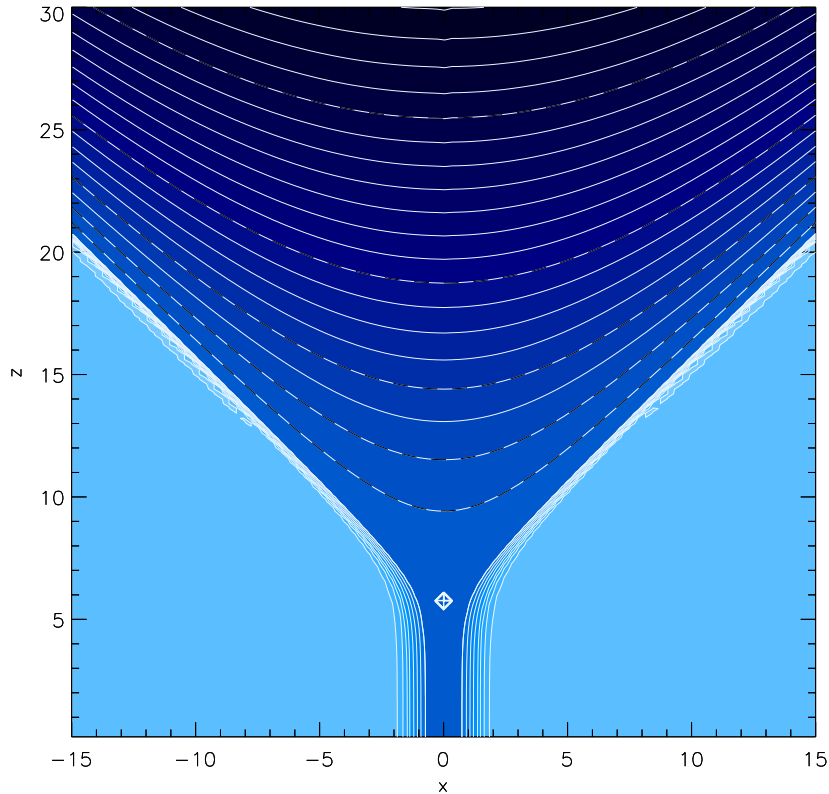}
\caption{An expanding bubble obtained by perturbing the initial data for a static O(3)-invariant solution. It can be seen that at late times, there are spacelike hyperbolic surfaces of homogeneity that fill the expanding bubble wall. The surfaces of constant field (black lines) are fit to hyperbolas (white-dashed lines), with origins indicated by the diamonds.
\label{fig-O(3)singlebubble}
}
\end{figure*}

We now consider more general O(3)-invariant initial data. Solving Eq.~\eqref{eq-O3initial} yields a static, spherically symmetric bubble configuration. The surfaces of constant field in this case would be purely timelike. This static solution is, however, unstable to either collapse or expansion under small perturbations. Perturbing the static solution such that an expanding solution eventually results, we obtain the evolution in Fig.~\ref{fig-O(3)singlebubble}. As the initial configuration evolves, the bubble wall begins to accelerate due to the pressure gradient across the wall. As one might expect, at late times memory of the initial conditions fades, and this acceleration becomes time-independent, yielding a hyperbolic trajectory for the wall. As this occurs, the surfaces of constant field inside the bubble go from being purely timelike to purely spacelike! Because this region is bounded by timelike hyperbolas composing the constantly accelerating wall, we might expect that SO(3,1) symmetry arises at late times.

Indeed, we find this to be the case. In Fig.~\ref{fig-O(3)singlebubble}, we have fit hyperbolas (white dashed lines) to the constant-field surfaces (black lines); diamonds mark the hyperbolas' origins. It can be seen that the hyperbolas are an excellent fit; moreover, even the earliest spacelike surfaces are hyperbolic, and their origins all lie at a single point at the apex of a lightcone of constant field. While the two latter points are specific to the static O(3) bubble, the result of spacelike hyperbolas is not: experimentation with a variety of O(3) symmetric `lumps' that lead to an expanding interface indicates that the generation of hyperbolic sections at late times and at large radius is very generic. Thus, we see that infinite, spacelike surfaces of homogeneity can spontaneously arise during field evolution from asymmetric initial data.

Because the interior of the bubble does not initially have a nice foliation into homogeneous spacelike hypersurfaces, the question of inflation arising inside the bubble is somewhat subtle. Applying Eq.~\eqref{eq-SECfinal} to every point in the simulation grid, we find that the example in Fig.~\ref{fig-O(3)singlebubble} violates the SEC everywhere, a requirement for inflation discussed in Sec.~\ref{sec-method}. Inside the bubble, for our assumed potential the field velocity is simply not large enough to overwhelm the large value of the potential; the surfaces of constant field approach homogeneous spatial slices, and eventually the field relaxes to the true vacuum where reheating presumably occurs. Thus, a phenomenologically viable epoch of inflation occurs inside the bubble.

At early times in the evolution, asymmetric initial data will define a preferred center of the bubble. However, the residue of this preferred frame disappears very rapidly in  the trials we have simulated. Therefore, we expect that models where inflation occurs from a spherically symmetric expanding field configuration will be indistinguishable from open inflation at late times. (It is plausible that the condition of spherical symmetry can be relaxed as well, since asphericities on an expanding bubble wall shrink~\cite{Adams:1989su}.)

\subsection{Bubble collision simulations}\label{sec-collsimulations}

Let us now turn to simulation of collisions between two bubbles. The hyperbolic symmetry present in the collision of two bubbles allows us to evolve the field using Eq.~\eqref{eq-hypevolve}, and we will use as initial data the bubble configurations found from Eq.~\eqref{eq-CDLinitialdata}, as discussed in Sec.~\ref{sec-method}.

\begin{figure*}[htb]
\includegraphics[width=12cm]{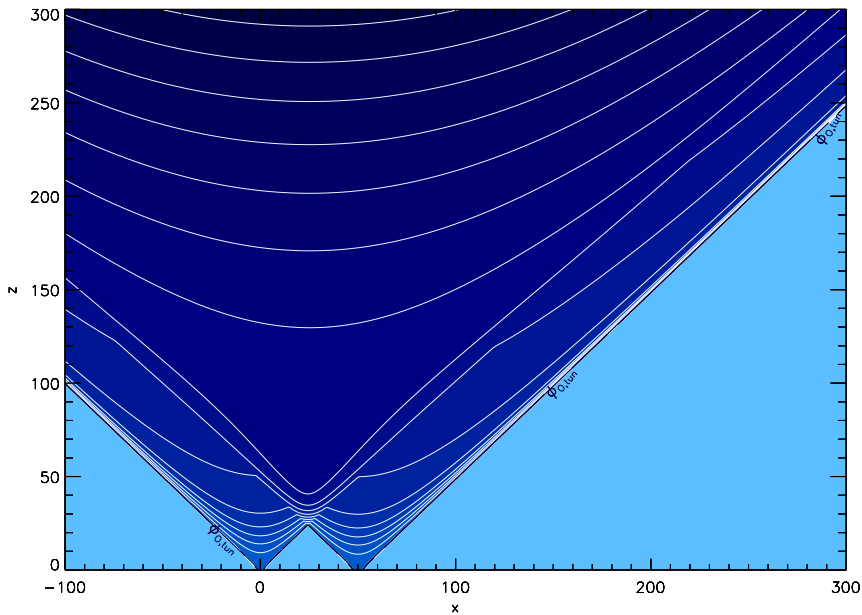}
\caption{The collision of two identical bubbles. Each point in the simulation plane corresponds to a 2-hyperbola of radius $z$. Numerical fitting of surfaces that look hyperboloidal shows that in general, they actually are hyperboloids to excellent accuracy.
\label{fig-collision1}
}
\end{figure*}

We begin by simulating the collision between two identical bubbles. Using the fiducial potential, we evolve two identical copies of the initial data for the observation bubble. The evolution of one such bubble is shown in Fig.~\ref{fig-O(3,1)singlebubble}; the simulated collision is shown in Fig.~\ref{fig-collision1}. The disturbance due to the collision propagates along left and right directed (approximately) null rays emanating from the position where the two bubbles meet; this carries the energy of the colliding walls. Apparently, at late times the constant-field surfaces inside the region bounded by this disturbance are hyperbolas! Thus, in a way very reminiscent of the O(3)-invariant single bubble simulations of Sec.~\ref{sec-singlebubblesim}, a set of infinite spacelike surfaces of homogeneity spontaneously evolves. Fitting hyperbolas to surfaces of constant field, we find that the fits are very good, but that the origins of the hyperbolas coincide only at late times after the effects of the preferred frame introduced by the collision have faded.

\begin{figure*}[htb]
\includegraphics[width=12cm]{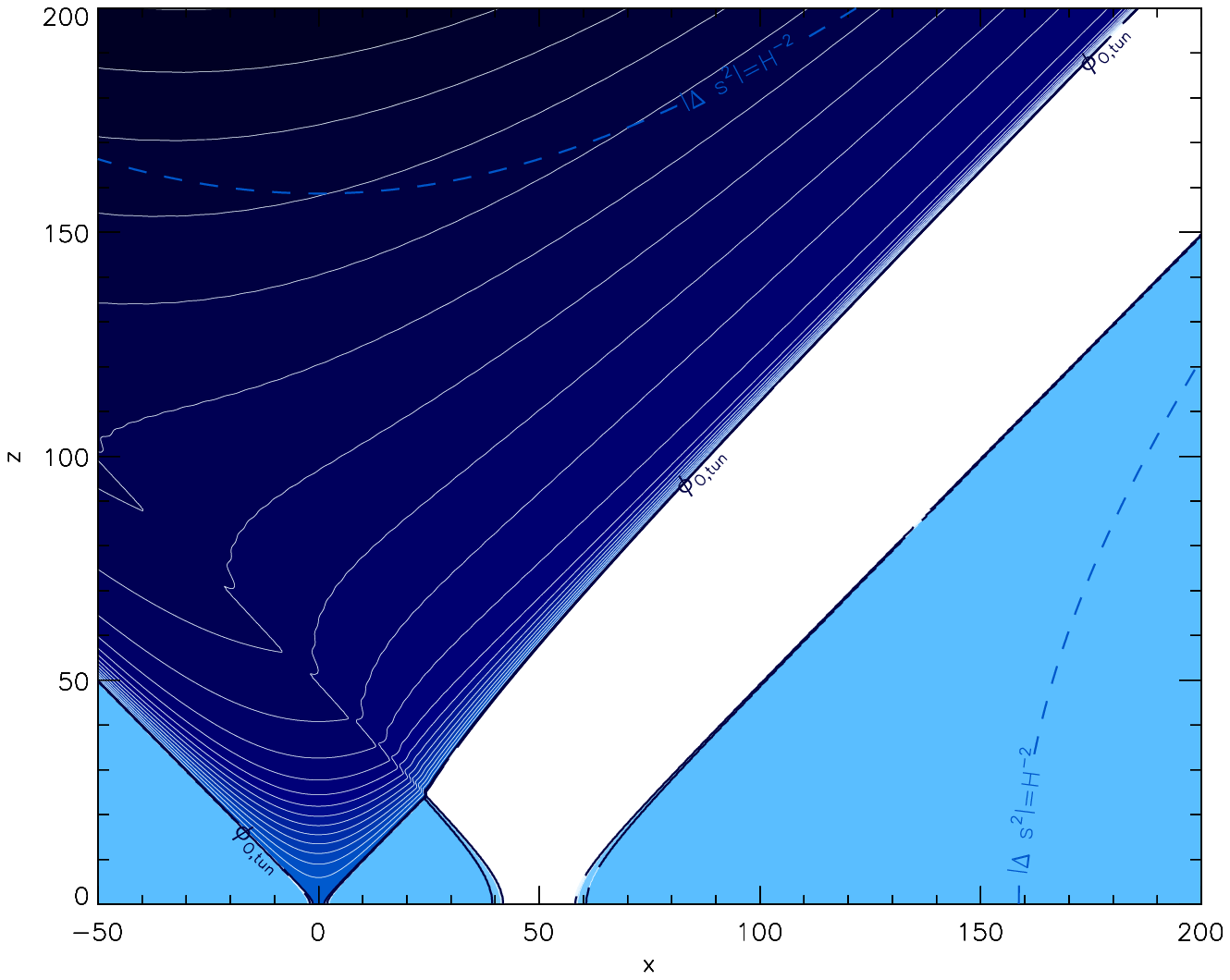}
\caption{
Collision of two bubbles.  Each point corresponds to a 2-hyperbola of radius $z$. The `observation' bubble is centered on $x=0$; it interpolates between the false minimum $\phi_{\rm F,min}$ and $\phi_{\rm O,tun}$ across the right barrier (see Fig.~\ref{fig-simpotential}). The `collision' bubble is centered on $x=50$; it interpolates between the false minimum and $\phi_{\rm C,tun}$ across the left barrier. The invariant distance from the origin, $\Delta s2 = H^{-2}$ which corresponds to the false-vacuum Hubble radius, is depicted; we expect that gravitational effects will be negligible within this region.
\label{fig-collision2}
}
\end{figure*}

We now move on to simulate collisions between two different types of bubbles. Again, using the fiducial potential, we evolve glued initial data for bubbles, interpolating from the false vacuum to each of the two other vacua. Fig.~\ref{fig-collision2} illustrates one such collision. Due to the relatively high energy of the colliding-bubble interior, a new domain wall separating the two different vacua forms and quickly accelerates away from the observation bubble. Additionally, a null disturbance propagates into the observation bubble from the location of the collision. (There is an accompanying disturbance inside the colliding bubble, but this lies in the near vicinity of the domain wall and cannot be seen in the plot.)

\begin{figure*}[htb]
\includegraphics[width=12cm]{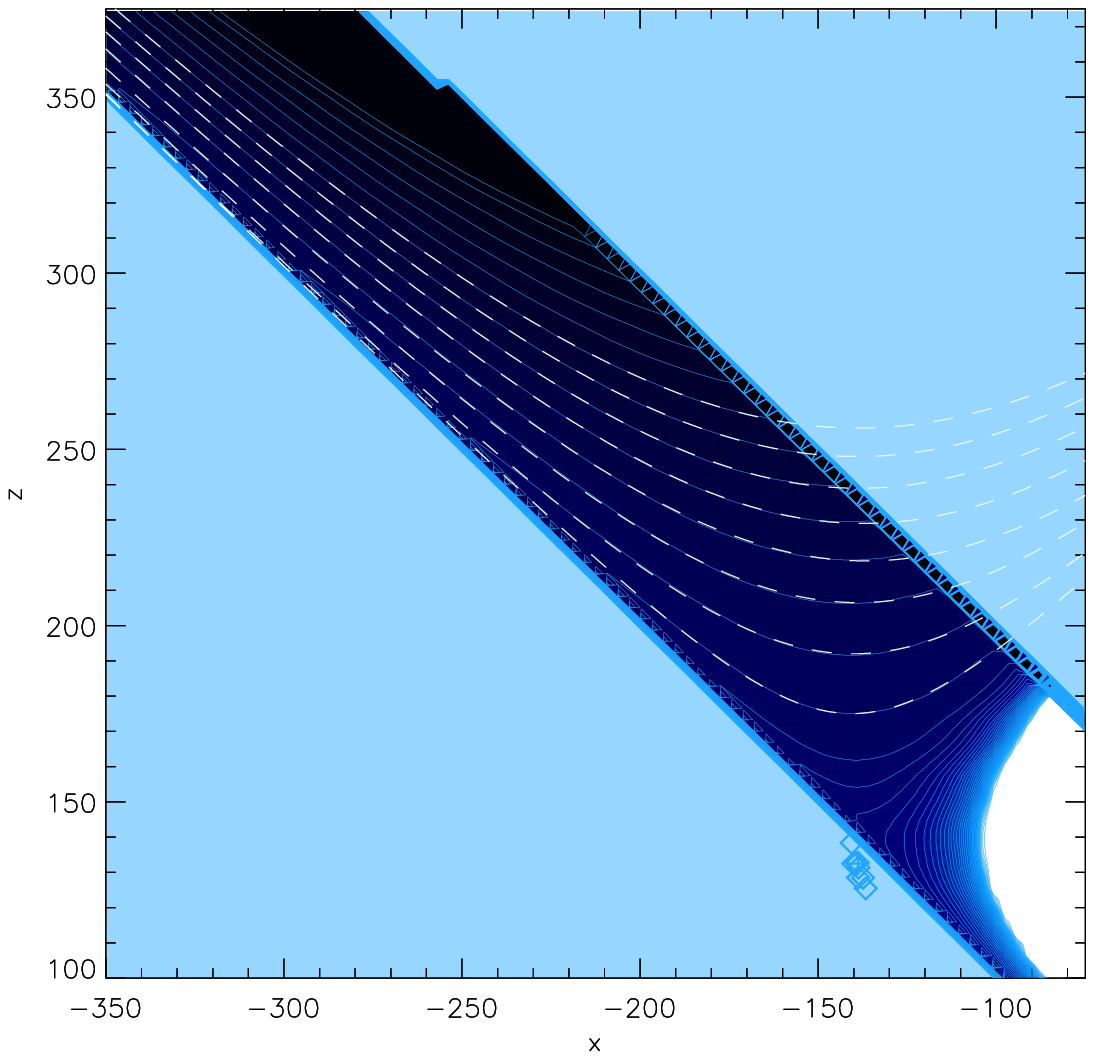}
\caption{
The same collision shown in Fig.~\ref{fig-collision2}, but boosted by $\gamma=8$ in the $-x$-direction and showing only the $(x-z)$ plane (see text).  The boundaries of the region correspond to the (highly boosted) edges of the simulation in the original frame, and the region shown corresponds to a region very near the domain wall in Fig.~\ref{fig-collision2}.
As in Fig.~\ref{fig-O(3)singlebubble}, the surfaces of constant field (black lines) are fit to hyperbolas (white-dashed lines), with origins indicated by the diamonds.  The (small) scatter in these origins is caused by the lower resolution available in this highly-boosted frame.
\label{fig-collision3}
}
\end{figure*}

The field falls away from the constantly accelerating post-collision domain wall, and at large $z$ appear to approach hyperbolic surfaces of homogeneity. As before, we can carefully fit hyperbolas to the surfaces of constant field, but now it is convenient to do so in a different Lorentz frame to more clearly show the surfaces closer to the post-collision domain wall. Boosting (with $\gamma=8$) in the $x$-direction, we obtain Fig.~\ref{fig-collision3}. (Such a boost would actually look awkward in the $(z,x,\chi,\varphi)$ coordinates, but in the special case of the $z-x$ plane these reduce to Minkowski $(t,x)$ coordinates in which the boost has a simple form.) It can be seen in Fig.~\ref{fig-collision3} that the hyperbolas fit the surfaces very well, and that origins of these hyperbolas (shown as diamonds) coincide in the boosted frame.\footnote{The origin of the hyperbola in the original frame can then be obtained by boosting back, though this turns out to exponentially magnify the uncertainty in the determination of the origin.}  Because such a boost transforms hyperbolas into hyperbolas of the same curvature (but different origins), the equal-field surfaces must be hyperbolas in (the $z-x$ plane of) the original frame also.

Extrapolating our results outside of the simulation region, the existence of hyperbolic surfaces of constant field lead us to conclude that regions to the future of a collision can possess infinite spacelike surfaces of homogeneity.\footnote{Here and elsewhere we have assumed that we can trust our non-gravitational solution to characterize the true gravitational solution far `up the lightcone' (at $z \simeq x\rightarrow\infty$). This is clear for an undisturbed CDL bubble, in which the solution depends only on the (dS or Minkowski) invariant distance from the nucleation point.  For interacting bubbles, especially given the tendency of the dynamics to restore SO(3,1) boost symmetry, we can imagine an invariant distance measure in which points `up the lightcone' are within a small invariant distance; this is arguably more relevant than any other distance measure in assessing the reliability of our flat-space results. Nonetheless, rigorously showing that our results hold far from the nucleation point would require running simulations with gravity.} Ultimately, this is a consequence of the constant acceleration of the post--collision domain wall, and therefore we expect this result to persist when gravitational back--reaction is included.

An analytical treatment of collisions including gravitational effects was presented in Paper II (see also~\cite{Wu:1984eda,Bousso:2006ge,Chang:2007eq}), where it was assumed that the outgoing null disturbances and the post-collision domain wall can be treated as infinitesimally thin shells. In addition to the back--reaction of the field energy, there will be a Hyperbolic Schwarzschild mass parameter, $M_\mathrm{obs}$, associated with the energy released during the collision. When the post-collision domain wall has a tension much smaller than the false vacuum Hubble scale and the bubble walls are nearly null at the time of collision (which is true in our simulations) the horizon radius associated with the Hyperbolic Schwarzschild space is given by
\begin{equation}
G M_\mathrm{obs} \simeq \frac{\left( H_F^2 z_c^2 - H_c^2 z_c^2\right) \left(1+ H_\mathrm{obs}^2 z_c^2 \right)}{2 \left(1+ H_F^2 z_c^2 \right)} \, z_c \,,
\end{equation}
where $H_F$ is the Hubble constant in the false vacuum, $H_c$ that of the colliding bubble's vacuum, $H_\mathrm{obs}$ that of the observation bubble evaluated on the other side of the post-collision domain wall, and $z_c$ the value of $z$ at the location of the collision. Since we consider positions $z_c \ll H_{F,c,\mathrm{obs}}^{-1}$, we will have $G M_\mathrm{obs} \ll z_c$, and gravitational effects should therefore be sub-dominant in the collisions that we consider.

Within this setup, it is also possible to model the back--reaction of the field energy. Following Ref.~\cite{Chang:2008gj}, one can introduce a test field (representing a slowly rolling inflaton) into the thin--wall collision spacetime. Imposing the boundary condition that this field is constant on the post--collision domain wall and that it matches onto the evolution of the field outside the future light cone of the collision, the surfaces of constant field in the collision region can be found. This analysis reveals, in agreement with our results in Minkowski space (and again due to the constant acceleration of the post--collision domain wall), that to the future of the collision event infinite spacelike surfaces of constant field result.

A hyperbola in the $x$--$z$ plane with an origin at $z=0$ corresponds to a locus of 2-hyperbolas (one for each value of $x$) that together form a 3-hyperbola. However, in our simulations, the hyperbolas that the constant-field surfaces approach at large $z$ will generally not have origins at $z=0$; this means that the 3-surface will {\em not} be a 3-hyperbola, but will possess an intrinsic anisotropy. To see this, assume the surfaces of constant field at large $x$ obey
\begin{equation}
z - z_0 = \left( \hat{\tau}^2 + (x - x_0 )^2 \right)^{1/2}
\end{equation}
where $\hat{\tau}$ labels the surfaces of constant field. In terms of the Minkowski coordinates, these surfaces are given by
\begin{equation}
t^2 = x^2 + w^2 + z_0 + \hat{\tau}^2 +2z_0 \sqrt{\hat{\tau}^2 + (x-x_0)^2}
\end{equation}
Further defining
\begin{equation}
x - x_0 = \hat{\tau} \sinh \hat{\xi}, \ \ z - z_0 = \hat{\tau} \cosh \hat{\xi} \,,
\end{equation}
the metric~\eqref{eq-xzmetric} becomes
\begin{equation}
ds^2 = - d \hat{\tau}^2 + \hat{\tau}^2 d \hat{\xi}^2 + \left( \hat{\tau} \cosh \hat{\xi} + z_0 \right)^2 \left( d\chi^2 + \sinh^2 \!\! \chi d\varphi^2  \right).
\end{equation}
The three-curvature scalar on surfaces of constant $\hat{\tau}$ is given by
\begin{equation}
R^{(3)} = - \frac{2 \cosh \hat{\xi} \left( 2 z_0 +3 \hat{\tau} \cosh \hat{\xi} \right)}{\hat{\tau} \left( z_0 + \hat{\tau} \cosh \hat{\xi} \right)^2},
\end{equation}
where the anisotropy is evident. Note that when $z_0 \rightarrow 0$, or if $\hat{\xi} \rightarrow \infty$ and/or $\hat{\tau} \rightarrow \infty$, the curvature is constant and given by $R^{(3)} = - 6 \hat{\tau}^{-2}$, recovering the Milne patch of Minkowski space (the open patch~\eqref{eq-frwpatch} with $a = \tau$) when $z_0=0$. Therefore, observers at large $\hat{\xi}$ for any $\hat{\tau}$ effectively reside in a universe where the original $SO(3,1)$ symmetry of an undisturbed bubble universe is restored. We have been unable to find a corresponding metric ansatz including gravitational back-reaction, but again expect this result to generalize.

\begin{figure*}[htb]
\includegraphics[width=12cm]{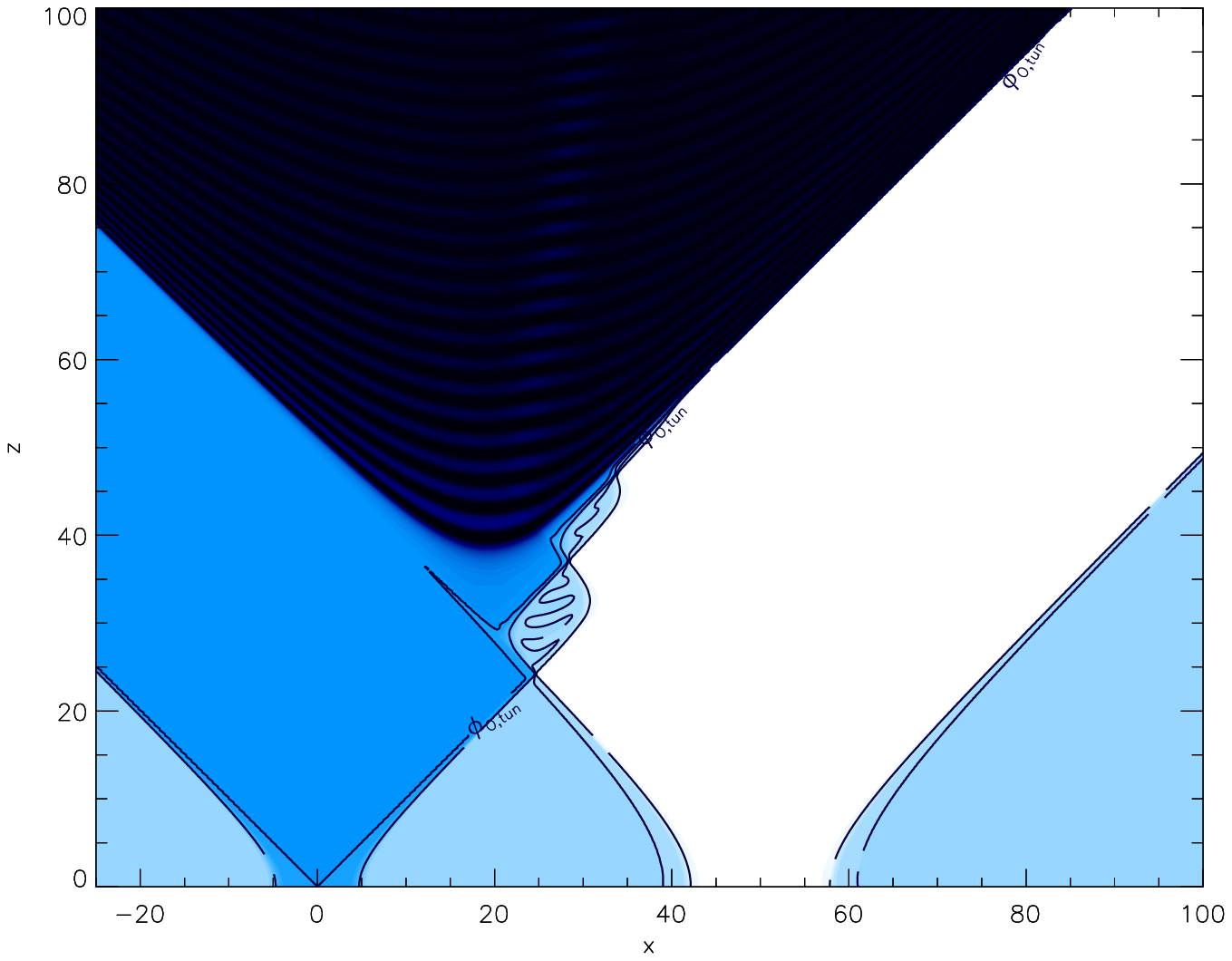}
\caption{
A collisions between bubbles in an `accidental' inflation scenario. Prior to the collision, the field is essentially homogeneous inside the observation bubble; but in the region hit by the bubble, the inflection point no longer coincides with zero kinetic energy in the field, and inflation is lost.
\label{fig-collision4}
}
\end{figure*}

We now address the question of inflation occurring to the future of a bubble collision. We consider two representative models for the epoch of inflation inside of the observation bubble: the fiducial large--field model used above, and the small--field `accidental' open inflation model described in Appendix~\ref{sec-potentials} and shown in Figure~\ref{fig-collision4}. Beginning with the large--field model, for the collisions depicted in Figs.~\ref{fig-collision1} and~\ref{fig-collision2}, the equal-field surfaces to the future of the collision are in the inflationary region of the potential, the SEC is violated, and the evolution is on small enough scales that it should still be well-described by the non-gravitational equations of motion we employ. A collision in the small field model is shown in Fig.~\ref{fig-collision4}. To the future of the collision, the field is immediately displaced from the inflationary region of the potential, and oscillates around the true vacuum minimum.

The origin of this striking difference is clear. The collision injects gradient and kinetic energy into the field. In the large--field models, this energy can be damped before the field overshoots the inflationary region of the potential. Including gravitational effects would seem to only help in this regard, since the attractor behavior due to Hubble damping (which is not accounted for in our simulations) will allow for a wide variety of initial conditions for the field. Small--field models are much more sensitive to initial conditions because the slow--roll parameters are satisfied only near an extremal point (see Appendix~\ref{sec-potentials} for further discussion), and the injected energy due to the collision causes the field to overshoot this region, spoiling inflation to the future of the collision. These results suggest that inflation can robustly occur to the future of a collision only in large--field models.

While our results appear quite general throughout the trails we have run, the details of the collisions depend, of course, on both the scalar potential and on the initial bubble separation. For example, increasing the initial separation produces a disturbance propagating into the observation bubble that is stronger and sharper (this is a kinematical effect due to the larger center of mass energy of the collision). Altering the potential so that the acceleration toward the colliding bubble is higher has a similar effect of creating a strong pulse of energy into the observation bubble. This pulse can even effectively create equal-field surfaces perpendicular to the observation bubble wall, which drive `plane waves' that propagate in the $+x$ direction.  (Even in this case we find that for large enough $z \simeq x$, the field eventually settles into spacelike hyperbolas.) This may have interesting implications for the properties of inflation to the future of the collision. We postpone a more detailed study for future work.

In summary, our numerical simulations indicate that infinite spacelike surfaces of homogeneity can exist to the future of a collision event.  Assuming our flat-space results generalize to the case with gravity included, far ``up the domain wall" from the collision, the degree of anisotropy on such surfaces decreases, and eventually the SO(3,1) symmetry of an undisturbed bubble is restored to great accuracy (although with a displaced origin). Thus, it is entirely plausible that our observable universe could exist to the future of a collision. However, we have found that inflation can occur in the post-collision region only if the potential is not of the small-field type.

\section{Implications}
\subsection{Where are the observers?}
\label{sec-obsloc}

Using the results of the previous section, we can now address the question of where observers are located inside of the bubble. As we have emphasized, bubble collisions will appear to have drastically different effects depending on the chosen congruence of observers. In addition, various members of a given congruence will detect different signatures of bubble collisions. In assessing the potential for observing collisions in our universe, it is therefore crucial to assess which phenomena are generic versus those seen only from very atypical locations.

Given a theory entailing bubbles with different properties, it is a major open question how to generate the probabilities governing what we should expect to observe. A popular and plausible approach is to compare relative frequencies of `observers', as estimated using some proxy such as number of galaxies, or physical volume on the inflationary reheating surface (see, e.g.,~\cite{Aguirre:2005cj,Winitzki:2006rn,Vilenkin:2006xv} for some discussion.) Aside from ambiguities in the choice of proxy, the difficulty confronting this program is that the volumes (or numbers) are infinite and there is no unique scheme for regularizing them.

In the absence of a complete measure, we might hope that concentrating on a single bubble will yield meaningful information about what we expect typical observers to experience.\footnote{However, the full measure may play a role in answering this question, see ~\cite{Bousso:2008hz,DeSimone:2008bq,DeSimone:2008if}.} In this case, the ambiguities described above might be substantially mitigated because (a) considering collisions there is a well-defined ``center" to the bubble~\cite{Garriga:2006hw}, (b) there can be a symmetry, of approximate uniformity along a spacelike hypersurface of constant-field, and (c) in this case physical volume on the post-inflationary reheating surface provides a compelling proxy for number of observers. Thus a natural way of comparing relative observer frequencies is to compute relative frequencies of physical volume on a surface of constant field within some radial distance from the bubble center, then send that radius to infinity. In the case of the relative frequencies of points with different collision events to their past, this can be done by calculating, as a function of a point's radius, the 4-volume in the past lightcone of that point available to nucleate bubbles.

This program was implemented in Ref.~\cite{Garriga:2006hw} and Papers I \& II, where it was assumed that (a) all bubbles nucleate out of the same false vacuum and (b) incoming bubbles produce no effect on the geometry of the observation bubble, so that the original open FRW foliation inside of the bubble can be used. In this case, the relevant past 4-volume scales, at large $\xi$, as $V_4\sim A\xi$, where the constant $A$ depends upon geometrical factors and on the cosmology inside the observation bubble.  At large $\xi$, this corresponds to a probability $1-e^{-\lambda A\xi}$ of being to the future of a collision, where $\lambda$ is the nucleation rate per unit 4-volume.  By the prescription above, this indicates that essentially {\em all} observers have at least one collision of this type to the past.

This effect can be understood in several ways.  First, an observer at $\xi\rightarrow\infty$ follows a worldline that looks null with respect to the `steady state' background frame defined by the bubble distribution; the high rate of incoming bubbles can then be attributed to a time dilation. Second, for fixed $\tau$ and $\xi\rightarrow\infty$, the observer position is arbitrarily far `up the lightcone' from the nucleation point. Thus there is an arbitrarily large 4-volume in the observer's past lightcone that exists outside the observation bubble wall. (On a conformal diagram, this looks like the inclusion of a point of future infinity in the observer's past lightcone.)  Third, in sufficiently simple spacetimes a global `boost' into the observer's reference frame can be defined, and performed without altering the physics. In this new frame, the region from which bubbles can nucleate in the observer's past lightcone extends to {\em past} infinity in a way that includes infinite 4-volume. (As shown in~\cite{Aguirre:2007an}, the area to the future of these collisions would cover nearly the full sky of an observer; this is best seen in the third view, in the frame of which these bubbles approach `from past infinity', and look as if they pass over nearly the whole observation bubble.)

What impact will including the effects of collisions on the bubble interior have on this picture? For a single collision, the results of the previous section indicate that the post-collision constant-field surfaces can asymptotically resemble a `displaced' version of those in the undisturbed bubble interior with some strongly affected interpolating region. In the asymptotic region, the bubble structure is essentially identical to that of the original bubble without collisions. However, this region will not generally be protected from additional collisions arising due to the nucleation of bubbles within bubbles or additional bubbles nucleating out of the false vacuum. In an undisturbed bubble, the typical distance in $\xi$ along a constant $\tau$ surface to a collision with a bubble that nucleates with rate $\lambda$ is~\cite{Garriga:2006hw} $\Delta \xi \sim \ln (1/\lambda)$. To be to the future of the first collision, an observer must (for the typically small values of $\lambda$) be at relatively large $\xi$. To the future of this first collision, most of the 4--volume to the past of a given point is in the direction of the original collision,\footnote{This is most easily seen in the (boosted) `observation' frame, although it is unclear how to perform this operation explicitly without embedding the spacetime in Minkowski space. For general collision spacetimes, this almost certainly cannot be done in 4+1D.} and one therefore expects additional collisions to come from essentially the same direction as the original. Because the region to the future of the first collision is similar to that of the undisturbed bubble, we can apply the above argument again to find the distance to the second collision and so on.

In this picture, it seems reasonable to use the prescription described above to assess what typical observers see. Consider a particular constant-field surface, and remove from it regions to the future of collisions in which the domain wall does {\em not} accelerate away, or that prevent inflation from occurring to their future (as per the `accidental inflation' case we exhibited in Sec.~\ref{sec-collsimulations}). In what is left, proceeding along the surface we encounter a number of `disturbed regions' caused by incoming bubbles, far away from which uniformity is restored. From the analysis of Ref.~\cite{Garriga:2006hw} and Paper I, we expect most observers to be to the future of at least one collision, where the results of the present paper have shown it is plausible for them to exist. (In fact, If observers can exist to the future of many collisions, then typical observers would seem to have an infinite number of collisions to their past, just as in the analysis of an undisturbed bubble.) However, in the regime of tiny nucleation rates the distance between subsequent disturbed regions will be very large -- because the typical distance between disturbed region is set by the comoving distance $\Delta \xi \sim \ln (1/\lambda)$ -- and therefore typical observers will exist far from such regions. We now discuss a few implications of this picture in more detail.

\subsection{Implications for observing collisions}
\label{sec-implications_obs}

In Sec.~\ref{sec-collsimulations} we argued that, far from a collision, the bubble interior spacetime approaches the SO(3,1) symmetry of an undisturbed bubble with homogeneous constant-$\tau$ surfaces. This suggests that observers in this region would see local homogeneity and isotropy. However, the structure of the spacetime indicates that the collision region (i.e. the region in which local homogeneity has {\em not} been restored) remains in the past lightcone of such observers, so we may ask whether the observers might see an effect of the collision on a spatial slice far in their past. If the observers are spatially far away from the collision, which is expected based on the arguments of the previous section, one might guess that the angular size of such an effect is small. Indeed, this is the case, which can be seen as follows.

Assuming local restoration of SO(3,1) symmetry, let us define a set of SO(3,1) compatible coordinates $(\tau,\xi,\theta,\varphi)$ (as in Eq.~\eqref{eq-OpenCoordTransform}) surrounding the observer at $(\tau_\mathrm{obs},\xi_\mathrm{obs})$, bounded by a `wall' indicating the edge of the region in which SO(3,1) has been restored. For a Minkowski background these coordinates are related to the Cartesian ones via
\begin{eqnarray}  \label{eq-OpenCoordTransform}
&& \hspace{-8pt} t = \tau \cosh\xi,\qquad\qquad\quad\;\,
   x = \tau \sinh\xi \cos\theta, \\
&& \hspace{-8pt} y = \tau \sinh\xi \sin\theta \cos\varphi \,, \ \
   w = \tau \sinh\xi \sin\theta \sin\varphi \,,  \nonumber
\end{eqnarray}
corresponding to the metric~\eqref{eq-frwpatch} with $a(\tau) = \tau$. We can also cover this region by the $(z,x,\chi,\varphi)$ coordinates of Eq.~\eqref{eq-HyperbolicCoordTransform}, and due to the SO(2,1) symmetry everywhere, and can usefully model the boundary by some function $x(z)$.

The past lightcone of the observer intersects a surface of $\tau=\tau_{\rm sky}$ in a 2-sphere, and the intersection of this 2-sphere with the wall $x(z)$ (if a solution exists) is a circle. To find the angular size $\psi$ of this circle, it is useful to boost to a frame in which the $\xi_{\rm obs}=0$ (this does not affect $\tau_{\rm obs}$ or $\tau_{\rm sky}$); in this frame $\psi = 2\theta_{\rm disk}$, where $\theta_{\rm disk}$ is given by the solving the wall-sky intersection.

To compute the intersection, we can write the past lightcone as $\xi=\xi_{\rm PLC}(\tau,\tau_{\rm obs})$, so $\tau=\tau_{\rm sky}$, $\xi=\xi_{\rm PLC}(\tau_{\rm sky},\tau_{\rm obs})$ defines the 2-sphere of the sky. Then, from the relationships~\eqref{eq-OpenCoordTransform} and \eqref{eq-HyperbolicCoordTransform} we obtain
\begin{eqnarray}
    \label{EqnTauExpr}
  \tau^2 &=& z^2 - x^2  \\
    \label{EqnXiExpr}
  \tanh^2\xi &=& \frac{x^2}{z^2} + \tanh^2\chi  \\
    \label{EqnThetaExpr}
  \tan^2\theta &=& \frac{z^2}{x^2} \, \sinh^2\chi
  \,.
\end{eqnarray}

These define the problem in general, but are complicated to solve due to the effect of the boost on the $x(z)$ surface (which turns it into a surface $x(z,\chi)$).  However, in the case $\xi_{\rm obs}\rightarrow \infty$ (in the unboosted frame), the problem simplifies.   First, we note that for radial rays $ds=0$ gives $\Delta\xi\equiv \xi_{\rm obs}-\xi_{\rm PLC} = \ln (\tau_{\rm obs}/\tau_{\rm sky})$, independent of the boost. Thus if $\xi_{\rm obs}\rightarrow\infty$ in the unboosted frame, then at the intersection of the sky and the wall, either $\Delta\xi \rightarrow \infty$, in which case $\tau_{\rm sky}\rightarrow 0$ (by the preceding equation), or $\xi_{\rm PLC} \rightarrow \infty$, in which case $\tau_{\rm sky}\rightarrow 0$ (by the coordinate relation $x=\tau\sinh\xi\cos\theta$).

But if $\tau_{\rm sky}\rightarrow 0$, then by Eq.~\eqref{EqnTauExpr} $x/z \rightarrow 1$, which by Eq.~\eqref{EqnXiExpr} gives $\chi\rightarrow 0$; putting these into Eq.~\eqref{EqnThetaExpr} yields $\theta\rightarrow 0$. That is, we expect that observers at large $\xi_{\rm obs}$ do not see the collision event unless they look to very early times, and even then the angle is negligibly small. This result is independent of the specific form of the trajectory $x(z)$ as long as it is not null and towards the observer (in which case our assumption of finite $x$ at the collision would not hold).
Another way to look at this is that by boosting the observer from large $\xi$ to the origin, we boost $x(z)$ to a null surface heading away from $\xi=0$; in Paper I we showed that in this case the area on the sky unaffected by the bubble collision, which in the present case is the area of interest, has vanishing angular size (further, in this computation the angular scale of the unaffected region went to zero very rapidly with increasing $\xi$).  

Unfortunately, when coupled with the discussion of Sec.~\ref{sec-obsloc}, this result suggests that most observers would see bubbles in their past that have tiny angular scale.

\subsection{Implications for eternal inflation measures}
\label{sec-implications_ei}

As mentioned above, eternal inflation faces a severe `measure problem' when computations of the relative frequencies of different classes of observers are desired.  If, for example, we wish to ask how many observers should find themselves in a bubble of `type $A$'  versus one of `type $B$', we must compare observer frequencies between bubbles even though (a) there are an infinite number of each type that nucleate, and (b) each bubble of each type is spatially infinite inside.
While there is no definitive method for making this comparison, it is quite possible that bubble collisions will play a key role insofar as the measure seeks to compare, say, the relative volumes on the reheating surfaces of two bubbles of types $A$ and $B$.

Because there is a center to each bubble defined by the lowest expected frequency of bubble impacts~\cite{Garriga:2006hw}, a plausible prescription would be to integrate radially from this center on an open slice $d\tau=0$ out to a comoving radius `cutoff' $\xi_{\rm cut}$ (for present purposes a cutoff in physical radius would act similarly).  For  a comoving cutoff this volume $V$ would scale exponentially with $\xi_{\rm cut}$, and scale with the total cosmic expansion between $\tau=0$ and reheating.

To include bubble collisions, consider the case where we have just the two bubble types $A$ and $B$. We first ask if the post--collision domain wall between bubbles $A$ and $B$ accelerates towards or away from $A$. If towards, then a piece of the reheating surface is removed. As per the discussion above in Sec.~\ref{sec-obsloc}, the fraction left {\em unremoved} (compared to the case of no collisions) is
$$
f\sim \exp\left( - A\lambda \xi_{\rm cut}\right),
$$
where $\lambda$ is the nucleation probability per unit four-volume. This vanishes as $\xi_{\rm cut}\rightarrow \infty$, though the total unremoved volume itself still diverges.

On the other hand, if the acceleration is away from $A$, then the results of this study suggest that the bubble may be seen as a small perturbation that partially disturbs, but does not remove, the reheating surface.\footnote{In the case of $A$-$A$ or $B$-$B$ collisions, things are less clear, since the bubbles `merge' as per the discussion in Sec.~\ref{sec-collsimulations}. In addition, for an unaccelerated wall between different vacua, the volume is arguably removed from both, as neither can support an infinite spacelike constant-field surface as in the accelerated case. See~\cite{Dahlen:2008rd} for further discussion of these cases.} The vanishing of the unaffected volume fraction implies if there were just two bubble types $A$ and $B$, the relative volumes in this prescription would depend profoundly on which way the $A$-$B$ domain wall accelerates.  What, then determines this?

Roughly, the domain wall accelerates away from the bubble with lower vacuum energy.  If both $A$ and $B$ have constant vacuum energy, the condition for the post--collision domain wall to accelerate away from $A$ is given in the thin--wall limit by (see Paper II)
\begin{equation}\label{eq:accelcondition}
H_B^2 - H_A^2 + k_{AB}^2 > 0 \,,
\end{equation}
where $H_{A,B}$ are the Hubble parameters for bubbles $A$ and $B$ and $k_{AB}$ is the tension of the post--collision domain wall. In our conventions, the square of the Hubble parameter will be positive for positive vacuum energies and negative for negative vacuum energies. For fixed tension, bubbles with a smaller vacuum energy will have fewer intruding domain walls and might be accorded a much higher measure by prescriptions like the one above.  This suggests the possibility of a non-anthropic explanation of low vacuum energy, as noted by~\cite{Chang:2007eq}. However, there is no inherent reason to prefer the lowest {\em positive} vacuum energy over the most negative one.

In addition, we are interested in bubbles that contain an epoch of inflation, during which the vacuum energy is positive and evolving. In this case, it is less clear that the asymptotic value of the cosmological constant alone determines the dynamics, and the thin--wall formula~\eqref{eq:accelcondition} will only be valid when $H_A$ is taken to be the inflationary Hubble parameter. Thus, in a model with a constant vacuum energy, the measure will tend to weight bubbles with the largest negative vacuum energy consistent with a chosen conditionalization, while in a model where bubbles contain a changing positive vacuum energy the measure would seem to favor a low energy scale of inflation. Other observables correlated with the propensity of the domain wall to accelerate toward or away could also have their measures affected.

In summary, the fact that the unaffected volume fraction on surfaces of constant $\tau$ inside any bubble goes to zero, together with our results that suggest infinite surfaces of homogeneity form to the future of collision events, make it plausible that the measure for eternal inflation would tend to reward observers that exist to the future of a collision inside of a bubble with very few intruding domain walls.

\section{Summary and discussion}
\label{sec-discuss}

A research program was outlined in Paper I to assess the possibility that a collision with another bubble `universe' might exist to our past and leave a direct observable signature of both eternal inflation and a complex inflaton landscape.  This would require that bubble collisions: (a) allow observers to their future; (b) exist to the past of a non-negligible fraction of observers so that they are not exceptionally rare; and (c) have effects that are actually observable.

In Paper I we argued (as had~\cite{Garriga:2006hw}) that in the limit where the bubble collision has {\em no} effect on the observer's bubble, essentially all observers should have collisions to their past that cover nearly the full sky, even when bubble nucleation rates are tiny; if nucleation rates are sufficiently large (which is not generally expected), observers could potentially see a small number of collision events covering a fraction of order unity of the sky. In Paper II we studied exact but idealized solutions for bubble collisions, which suggested that observers might survive to the future of even seemingly-severe bubble collisions; we also speculated how observable signatures of such collisions might appear (see also~\cite{Aguirre:2007gy,Chang:2007eq,Chang:2008gj}).

In this paper, we have analyzed in detail the structure of bubble collision regions to determine where physical observers could actually exist and what they might hope to see. Key to this discussion was defining `observers' inside the bubble universe as geodesics emanating from a spacelike reheating surface. In an undisturbed bubble universe, this regulates the perceived energy of collisions with other bubbles, and emphasizes the importance of understanding the surfaces of homogeneous field (which provide a natural time-slicing), including the effects of collisions.

To investigate the behavior of these surfaces of homogeneity, we have employed a set of numerical simulations of the evolution of scalar field configurations in flat spacetime. We derived a class of scalar potentials that could give rise to a phenomenologically viable model of Open Inflation, while allowing for a suitably small simulation region wherein gravitational effects of bubble solutions and their collisions could be neglected. In single bubble simulations, we have demonstrated that for spherically symmetric lumps of true vacuum, initially timelike surfaces of homogeneity spontaneously evolve into infinite, hyperbolic, spacelike surfaces of homogeneity. For observers far from the configuration's center these models should be indistinguishable from open inflation.

We have further shown that infinite hyperbolic surfaces of constant field can develop to the future of collision events, with complete homogeneity restored very far away from (but still to the future of) the region of initial impact. This indicates that inflation can produce nearly homogeneous and isotropic infinite reheating surfaces, and hence also our observable universe, to the future of a collision event. However, most observers (as measured by physical volume on the spacelike constant-field surfaces) in the post--collision region will perceive the angular scale of the collision boundary to be vanishingly small. This result was presaged in Paper I, which showed that such collisions cover the full sky; here we have further found that, except for a disk of infinitesimal angular size in some direction, a sky covered by a collision is indistinguishable from a sky unaffected by a collision.

In terms of the program outlined in Paper I, we are led to a picture in which (a) observers can exist to the future of bubble collisions as long as the collision wall accelerates away from the observers' bubble, (b) using the natural measure essentially {\em all} observers would have such collisions to their past, but (c) except for large nucleation rates or exceptionally rare observers, the observable {\em effect} of the collision is essentially invisible, sequestered to an unobservably tiny patch of the sky. We can hope that by chance, or due to some consideration not captured by the present analysis, or with sufficiently high nucleation rates\footnote{See Paper II for an analysis of the necessary rates given a realistic cosmology inside the bubble.} we are close enough to a collision region to observe its signatures. Nonetheless, our analysis suggests a more probable scenario where tantalizing clues of all of the interesting dynamics of eternal inflation are up there in some particular direction on the sky, but remain forever hidden from us like the tiniest of needles in the cosmic haystack.

If this is true, then as some consolation, the existence of an infinite reheating surface to the future of a collision could mean that collisions play a crucial role in measures for eternal inflation. Given bubble types $A$ and $B$ that collide, with the interpolating domain wall accelerating into $A$, bubble $A$ would have a fraction {\em one} of its volume removed via collisions with $B$-bubbles, whereas $B$-bubbles could be argued to be essentially unaffected.  Depending upon the measure prescription this could accord much -- even infinitely -- larger weight to $A$ bubbles, and in general to any observable correlated with an interpolating domain wall that accelerates away.  Such a large factor holds out the possibility of generating sharp predictions, so perhaps in this sense bubble collisions may produce observable signatures after all.

\begin{acknowledgments}
The authors wish to thank T. Banks, S. Carroll, M. Kleban, and M. Salem for helpful discussions. MJ acknowledges support from the Moore Center for Theoretical Cosmology at Caltech. AA and MT were partially supported by a `Foundational Questions in Physics and Cosmology' grant from the Templeton  Foundation and NSF grant PHY-0757912 during the course of this work.
\end{acknowledgments}

\appendix

\section{Building models of open inflation}\label{sec-potentials}

Unlike in many inflation models, in open inflation the same scalar potential determines both the properties of and initial conditions for the inflaton dynamics. This property, along with the seeming ubiquity of first order phase transitions in models of high energy physics, has become a strong motivation for considering such models. In this appendix, we elaborate on these theoretical motivations and construct potentials used in the numerical simulations of Sec.~\ref{sec-simulationssetup}. We consider both `large-field models where the distance in field space traversed during inflation is greater than $M_{\rm pl}$, and `small-field' models in which it is not.

To begin our discussion, assume that the same scalar field is responsible for both tunneling and driving the epoch of inflation inside the bubble. In this case, there is a tension between guaranteeing slowÐroll and guaranteeing the existence of a CDL instanton, as first noted in Ref.~\cite{Linde:1998iw}. In order to find a CDL instanton, we must have
\begin{equation}
\frac{V''}{V} \agt 8 \pi
\end{equation}
near the instanton endpoints. (This condition is determined by comparing the size of the instanton to the Euclidean time taken to make the traverse between each side of the potential barrier.) However, the slowÐroll conditions
\begin{equation}
\epsilon \equiv  \left( \frac{V'}{V} \right)^2 \ll 16 \pi, \ \ \ \eta \equiv \frac{V''}{V} \ll 8 \pi
\end{equation}
require exactly the opposite inequality, implying that a successful single-field model of open inflation must possess a sharp feature across which the second derivative decreases significantly, while remaining flat enough to maintain slow-roll.

From the perspective of model building, this is rather unnatural since it requires a hierarchy in the scales over which the potential varies.  However, this does not preclude the construction of successful models of open inflation, since the form of the potential outside of the instanton endpoints, where slowÐroll is to occur, has no effect on the instanton itself.

In the literature, a number of assumptions are often made about the properties of vacuum bubbles in order to simplify the calculation of the nucleation rate or the form of the postÐnucleation spacetime. Specifically, it is desirable for bubbles to possess a wall whose thickness is much smaller than the radius at nucleation, which is in turn much smaller than the false vacuum de Sitter horizon size. This set of circumstances allows one to treat the wall as an infinitesimally thin membrane of fixed tension interpolating between the instanton endpoints (this is known as the `thin-wall' approximation), neglect gravitational effects in the computation of the instanton (we will refer to this as the `small--bubble' approximation), and approximate the bubble wall as a light cone in some situations. These requirements amount to additional restrictions on the form of the scalar potential comprising the bubble. Because the flat-space simulations of Sec.~\ref{sec-simulationssetup}, as well as the assumption of isolated pre-collision bubbles, make sense only in the small-bubble approximation, it is desirable to find a phenomenologically viable model for inflation in which this is valid.

A toy model for large-field open inflation was presented in~\cite{Linde:1998iw}. However, it is difficult to find parameters in this model where there is a thin-wall or small initial radius for the bubble, rendering this potential unsuitable for our simulations. In order to have parametric control over the thickness and initial radius of the bubble wall, while retaining a phenomenologically viable period of inflation, we construct a {\em piecewise} potential containing sufficiently many tunable parameters. While physically unmotivated, especially since we will only require continuity of the first and second derivatives, the generic shape of the potential is illustrative.

The potential, shown in Fig.~\ref{fig-simpotential}, consists of a total of four segments. We employ a potential with three vacua for the study of bubble collisions in Sec.~\ref{sec-collsimulations}, so there are two potential barriers: one for $\phi > 0$ and another $\phi < 0$. These barriers are constructed from the potential
\begin{equation}\label{eq:v1v2}
V_{1,2} = \mu_{1,2}^4 \left( \frac{(\phi-\hat{\phi}_{1,2})^2}{M_{1,2}^2} - 1 \right)^2 \pm \alpha_{1,2} \frac{(\phi-\hat{\phi}_{1,2})}{M_{1,2}} + C_{1,2}.
\end{equation}
As long as the overall height of the potential $C_{1,2}$ is not too large, and $M_{1,2} \ll M_{\rm pl}$, gravitational effects are not a dominant factor in the computation of the instanton. When $\alpha \ll 1$, the thin-wall approximation holds, and it is possible to calculate the tension of the bubble wall and its initial radius analytically~\cite{Coleman:1977py}:
\begin{equation}\label{eq-tensionR0}
\sigma = \frac{4}{3} M_{1,2} \mu_{1,2}^2, \ \ \ R_0 = \frac{3 \sigma}{\alpha} = \frac{4}{\alpha} \frac{M_{1,2}}{\mu_{1,2}^2}.
\end{equation}
When $\alpha \alt 1$, this equation will still be a good estimator for the bubble size. In order to neglect gravitational effects in the computation of the instanton, we require that the initial radius of the bubble is much smaller than the false vacuum de Sitter horizon size:
\begin{equation} \label{eq-deSitterHorizonSize}
H R_0 \simeq \frac{1}{\alpha} M_{1,2} \sqrt{\frac{V_0}{\mu_{1,2}^4}} \ll 1.
\end{equation}
It can be seen that by taking $M_{1,2} \ll 1$, one can make $H R_0$ arbitrarily small.

Numerically finding the instanton using Eq.~\eqref{eq-CDLinitialdata}, yields the instanton endpoints as well as the field configuration between them. At the true vacuum endpoint of $V_2$, we match onto a quadratic potential segment
\begin{equation}
V_3 = \frac{M_3^2}{2} (\phi-\hat{\phi}_{3})^2 + C_{3}
\end{equation}
that interpolates between the tunneling and inflationary parts of the potential. Specifying the length of this segment, and then requiring continuity of the potential and its first derivative both at the instanton endpoint and the junction to the next segment, determines the free parameters $M_3$, $\hat{\phi}_3$, and $C_3$.

The last segment is another quadratic potential that drives inflation:
\begin{equation}
V_4 = \frac{M_4^2}{2} (\phi-\hat{\phi}_{4})^2.
\end{equation}
The position in field space where the matching to the $V_3$ segment occurs determines the approximate position where inflation begins (approximate, because of the potentially steep interpolating segment $V_3$ the field must evolve over). In large-field models of inflation, the attractor behavior of the equations of motion will quickly bring the trajectory to the inflationary solution. In the presence of a quadratic potential, the inflationary parameters are given by (e.g.,~\cite{Liddle:2000cg})
\begin{equation}
N_e \simeq 2 \pi \phi_i^2, \ \ \frac{\delta \rho}{\rho} \simeq M_4 \phi^2, \ \ n_s \simeq .96
\end{equation}
where $N_e$ is the number of efolds with $\phi_i$ the displacement of the field from vacuum at the beginning of inflation; $\delta \rho /\/ \rho$ is the power in scalar modes with $\phi \sim \phi_i$ for a minimal number of efolds; and $n_s$ is the scalar spectral index evaluated at large scales. In order to get a minimal number of efolds, $N_e \sim 60$, we must take $\phi_i \sim 3$. This constrains where we match segments $V_3$ and $V_4$. To get the correct power in scalar modes $\delta \rho / \rho \sim 10^{-5}$, we must take $M_4 \sim 10^{-6}$. The spectral index of $n_s \simeq .96$ is in good agreement with the central value for WMAP 5-year data.

This set of models allows for a phenomenologically acceptable epoch of large-field inflation with an adjustable number of efolds. Further, the parameters in $V_{1,2}$ can be tuned to produce bubbles that are arbitrarily small compared to the false vacuum horizon size, and where the wall is arbitrarily thin. Thus, many of the simplifying assumptions made in the literature on eternal inflation and bubble collisions can be satisfied.

Small-field inflation near an approximate inflection point arises in a number of supergravity and stringy models of inflation, as has been explored in Refs.~\cite{Allahverdi:2006iq,Allahverdi:2006we,Baumann:2007np,Baumann:2007ah,Baumann:2008kq,Cline:2007pm,Panda:2007ie}, and has been dubbed 'accidental' inflation by Linde and Westphal~\cite{Linde:2007jn}. It is possible to construct phenomenologically viable models, with a variable number of efolds. However, the initial condition for the field must lie very near the approximate inflection point in order to avoid over-shooting~\cite{Brustein:1992nk} the region where the slow-roll parameters are small. In addition, inflation occurs at a fairly low scale in these models (because of the smallness of the potential gradient, the Hubble scale must be rather low to prevent too much power in scalar modes), and one can argue that it is difficult to arrange for such a (relatively) large patch being sufficiently homogeneous for inflation to begin.

Both problems can be largely alleviated in models of 'accidental open inflation,' in which inflation follows tunneling, and the instanton guarantees a homogeneous field configuration with zero kinetic energy arbitrarily close to the approximate inflection point.\footnote{The ability to alleviate the initial conditions problem in small-field models by invoking eternal inflation has also been discussed in~\cite{Linde:2007jn,Boubekeur:2005zm} in a slightly different context. In Ref.~\cite{Freivogel:2005vv} it was shown that the Hubble damping due to curvature largely alleviates the overshoot problem in open inflation. However, the number of efolds in models of accidental inflation is extremely sensitive to the initial position of the field~\cite{Baumann:2007np}, so this consideration probably does not suffice to solve the problem unless the instanton endpoint is close to the inflection point, as discussed here.} A similar observation was made by Cline~\cite{Cline:2005vg}, and is elaborated on in~\cite{Cline:2008fk} with an emphasis on the overshoot problem in models of D-brane inflation.

In single-field models of open inflation, even though satisfying $\eta \ll 1$ is impossible at the instanton endpoint, the requirement of a thin wall bubble helps to ensure that $\epsilon \ll 1$, making this a natural point to match onto a small-field inflationary potential. In order to produce a thin-wall bubble, the field must loiter near the endpoints of the instanton for a period of Euclidean time that is much longer than it spends traversing the potential barrier. This can only occur if both endpoints lie exponentially close to a critical point (or approximate critical point), where the gradient of the potential is extremely small, and therefore $\epsilon \ll 1$. The second derivative (of the lorentzian potential) must be positive at both endpoints for loitering to occur, and must be significant in magnitude, as discussed above. This imposes a strict condition on the form of the potential in the neighborhood of the true vacuum endpoint if we require that the potential is monotonically decreasing on this side of the barrier: the second derivative must very quickly decrease in magnitude, and the instanton endpoint will therefore lie in the near-neighborhood of an approximate inflection point, where both slow-roll conditions are satisfied! Therefore, a correlation will exist between thin-wall bubbles and the presence of an inflection point in the neighborhood of the instanton endpoint.

In order to implement such a small--field model in our simulations, we construct another piecewise potential. We use the two potential segments of Eq.~\eqref{eq:v1v2} to describe the potential barriers out of the false vacuum, allowing us to tune the initial radius of the bubbles. However, unlike in the large field model, we match onto a cubic potential of the form
\begin{equation}
V_{c} = \mu_c^4 \left( \frac{1}{12} - \frac{1}{3} \frac{\phi^3}{M_c^3} + \frac{1}{4} \frac{\phi^4}{M_c^4} \right)
\end{equation}
which has an inflection point at $\phi = 0$. The matching is performed between the minimum of the potential Eq.~\eqref{eq:v1v2} and the inflection point, producing the potential shown in Fig.~\ref{fig-simpotential2}. The field's initial condition is sufficiently close to the inflection point to produce a phenomenologically viable period of inflation, with a large number of efolds.

\begin{figure*}
\begin{center}
\includegraphics[height=6.5cm]{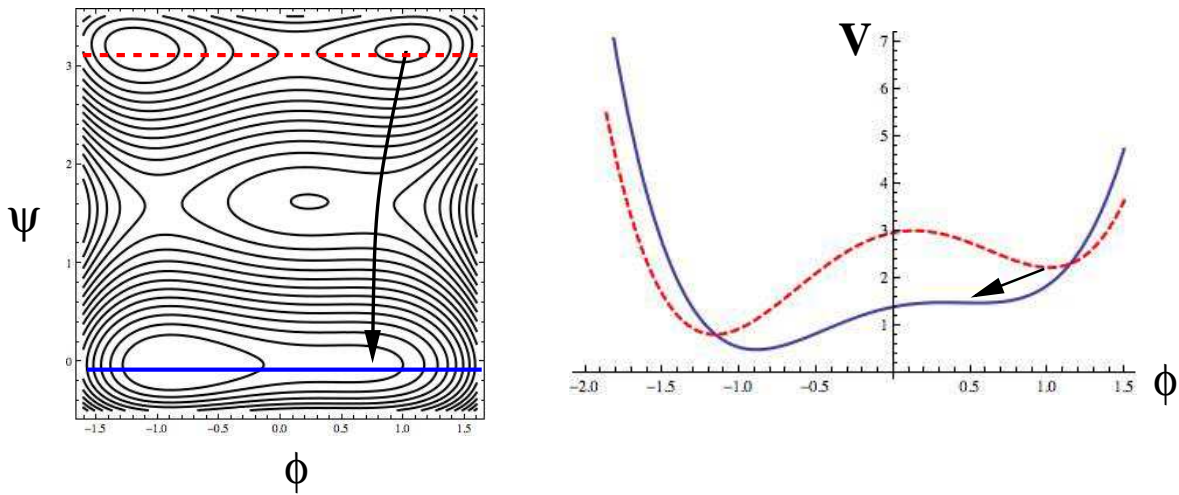}
\end{center}
\caption{A toy model for Accidental Open Inflation with $\{ v_0 = 1.36, \lambda_1 = .65, \lambda_2 = -1.2, \lambda_4 = 1, \epsilon_1 = .788, \epsilon_2 = 9.778, \epsilon_3 = -2 \pi, \epsilon_4 = 1, g= .12 \}$. The field tunnels in the $\psi$ direction, beginning an epoch of open inflation near the inflection point in the $\phi$-direction.}
\label{fig:toyaccident}
\end{figure*}

Models of multi-field accidental open inflation are also possible. For example, consider the toy model
\begin{eqnarray}
V  &=&  V_0 + \lambda_1 \phi + \lambda_2 \phi^2  + \lambda_4 \phi^4 \nonumber \\ & + & \epsilon_1 \psi + \epsilon_2 \psi^2 + \epsilon_3 \psi^3 + \epsilon_4 \psi^4 + g \phi^2 \psi^2 .
\end{eqnarray}
Turning off the coupling between the $\phi$ and $\psi$ sectors (sending $g \rightarrow 0$), we will choose parameters $\epsilon_{1,2,3,4}$ and $\lambda_{1,2,4}$ such that both the $\psi$ and $\phi$ sectors each have two minima and an intervening potential barrier. Turning the interaction back on, the mass term in each of the two sectors will be field-dependent. By tuning parameters, it is possible to construct a two field model of accidental open inflation where a tunneling event across a barrier in the $\psi$-direction places the field in the near vicinity of an inflection point in the $\phi$ direction, subsequently driving an epoch of inflation. An example of such a potential is shown in Fig.~\ref{fig:toyaccident}. In order for this model to be successful, the value of  $\psi$ in the vicinity of the instanton endpoint must be such that an inflection point or approximate inflection point exists in the $\phi$ direction. There will be an approximate inflection point when
\begin{equation}
\lambda_2 + g \psi^2 \agt - \frac{3}{2} \lambda_1^{2/3} \lambda_4^{1/3}
\end{equation}
with an exact inflection point for the equality. Crucial to this model is the ability of the VEV of $\psi$ to change finely enough during the tunneling event that a minimum in the $\phi$ direction becomes an approximate inflection point satisfying the above inequality. For an implementation of this model in the context of brane inflation, see~\cite{Cline:2005vg}.

\section{Strong energy condition for scalar fields}\label{sec-SEC}

In this appendix, we derive a useful formula for determining when the Strong Energy Condition (SEC) is violated by a scalar field configuration. The SEC is satisfied when
\begin{equation}\label{eq-SEC}
T_{\mu \nu} t^{\mu} t^{\nu} \geq \frac{1}{2} {T^{\lambda}}_{\lambda} t^{\sigma} t_{\sigma},
\end{equation}
for all timelike vectors $t^{\mu}$.

The energy momentum tensor for  a minimally coupled scalar field is given by
\begin{equation}
T_{\mu \nu} = \nabla_{\mu} \phi \nabla_{\nu} \phi - g_{\mu \nu} \left[ \frac{1}{2} g^{\rho \sigma} \nabla_{\rho} \phi \nabla_{\sigma} \phi + V(\phi) \right].
\end{equation}
Evaluating the left hand side of the inequality~\eqref{eq-SEC},
\begin{equation}
T_{\mu \nu} t^{\mu} t^{\nu} = \left( t^{\mu} \nabla_{\mu} \phi \right)^2 + \left( \frac{1}{2} g^{\rho \sigma} \nabla_{\rho} \phi \nabla_{\sigma} \phi + V(\phi) \right).
\end{equation}
Evaluating the right side of the inequality,
\begin{eqnarray}
\frac{1}{2} {T^{\lambda}}_{\lambda} t^{\sigma} t_{\sigma} &=& -\frac{1}{2} {T^{\lambda}}_{\lambda} \\ &=&  \frac{1}{2} g^{\rho \sigma} \nabla_{\rho} \phi \nabla_{\sigma} \phi + 2 V(\phi)
\end{eqnarray}

We can therefore write the SEC as
\begin{equation}
\left( t^{\mu} \nabla_{\mu} \phi \right)^2 - V \geq 0
\end{equation}
We would like to evaluate this condition at an arbitrary collection of points. To do so, we assume a set of timelike vectors at each point which have negative unit norm $t^{\mu} t_{\mu} = -1$ (essentially a four-velocity). We can go to locally inertial coordinates at each point, and the most stringent test of the bound will come from considering vectors pointing along the spatial gradient of the field. This defines a congruence that can be written in locally inertial coordinates as
\begin{equation}
t^{\mu} = (\gamma, \gamma v, 0, 0),
\end{equation}
where, $\gamma = (1 - v^2)^{-1/2}$.

Substituting for $t^{\mu}$ in the local inertial frame yields
\begin{equation}
\left( \gamma \partial_t \phi + \gamma v \partial_{\hat{n}} \phi \right)^2 - V \geq 0,
\end{equation}
where $\hat{n}$ indicates the direction along the gradient. We can substitute for $\gamma$ and then look for roots of the equality as a function of $v$:
\begin{eqnarray}
v &=& \frac{1}{(\partial_{\hat{n}} \phi)^2 + V(\phi)} \left[ - (\partial_{\hat{n}} \phi) (\partial_t \phi) \right. \nonumber \\ && \left. \pm \sqrt{V(\phi) \left[ (\partial_{\hat{n}} \phi)^2 - (\partial_t \phi)^2 + V(\phi) \right]} \right]
\end{eqnarray}
When there is a real root, then we will have a violation of the SEC at the point being evaluated. We can therefore re-state the SEC as
\begin{equation}\label{eq-SECfinal}
(\partial_{\hat{n}} \phi)^2 - (\partial_t \phi)^2 + V(\phi) < 0
\end{equation}
where these quantities are evaluated in {\em any} locally inertial frame (as one would expect from the SEC being a covariant condition).

\bibliography{bubbles3}

\end{document}